\newif\ifepjcclass
\IfFileExists{svjour3.cls}{\IfFileExists{svepjc3.clo}{\epjcclasstrue}{\epjcclassfalse}}{\epjcclassfalse}
\ifepjcclass
  \documentclass[twocolumn,epjc3]{svjour3}
  \journalname{Eur. Phys. J. C}
\else
  \documentclass[11pt,twocolumn]{article}
  \usepackage[a4paper,margin=1.8cm]{geometry}
\fi
\usepackage{amsmath,amssymb,bm,graphicx,booktabs}
\usepackage[T1]{fontenc}
\usepackage{microtype}
\usepackage{url}
\usepackage{cite}
\makeatletter\@ifundefined{ruledtabular}{\newenvironment{ruledtabular}{\begin{center}}{\end{center}}}{}\makeatother
\newcommand{\dd}{\mathrm{d}}
\newcommand{\ii}{\mathrm{i}}

\newcommand{\Boxg}{\Box}

\begin{document}
\title{Geometric criticality and regular quasibound spectra of axial Proca fields in Schwarzschild spacetime}
\ifepjcclass
\author{Antonio Sousa Ribeiro\thanksref{e1,addr1} \and Francisco Artur Pinheiro Alves J\'unior\thanksref{e2,addr2}}
\thankstext{e1}{e-mail: antonius.ribeirus@ifpi.edu.br}
\thankstext{e2}{e-mail: francisco.artur@univasf.edu.br}
\institute{Instituto Federal de Educa\c{c}\~ao, Ci\^encia e Tecnologia do Piau\'i (IFPI), Campus S\~ao Raimundo Nonato, 64770-000 S\~ao Raimundo Nonato, PI, Brazil\label{addr1} 
\and Universidade Federal do Vale do S\~ao Francisco (UNIVASF), Campus Serra da Capivara, 64770-000 S\~ao Raimundo Nonato, PI, Brazil\label{addr2}}

\maketitle
\begin{abstract}
We investigate how a local geometric critical point of the odd-parity Proca potential relates to the global quasibound spectrum in Schwarzschild spacetime. After normalization by $\mu^2$, the axial potential $V_{\rm ax}=(1-2M/r)[\ell(\ell+1)/r^2+\mu^2]$ has the same radial form as the timelike-geodesic effective potential. Its outer minimum and inner maximum merge at $r=6M$, the Schwarzschild innermost stable circular orbit, when $(M\mu_c)^2=\ell(\ell+1)/12$. With $\delta=1-(M\mu)^2/(M\mu_c)^2$, we derive exact fold scalings: stationary-point separation $\propto\delta^{1/2}$, barrier--well depth $\propto\delta^{3/2}$, and curvature magnitudes $\propto\delta^{1/2}$. Continued-fraction calculations for the first three axial quasibound branches with $\ell=1,2,3$, cross-checked for selected modes by independent Riccati shooting, show regular frequency continuation across the threshold and approximately linear near-critical behavior. At fixed radius the radial operator is affine in $\delta$, supporting a conditional simple-root analytic interpretation. The main result is the explicit distinction between geometric potential criticality and spectral degeneracy for the branches examined.
\end{abstract}
\else
\author{Antonio Sousa Ribeiro\\\small Instituto Federal do Piau\'i (IFPI), Campus S\~ao Raimundo Nonato\\\small antonius.ribeirus@ifpi.edu.br \and Francisco Artur Pinheiro Alves J\'unior\\\small Universidade Federal do Vale do S\~ao Francisco (UNIVASF)\\\small francisco.artur@univasf.edu.br}
\date{}
\maketitle
\begin{abstract}
We investigate how a local geometric critical point of the odd-parity Proca potential relates to the global quasibound spectrum in Schwarzschild spacetime. After normalization by $\mu^2$, the axial potential $V_{\rm ax}=(1-2M/r)[\ell(\ell+1)/r^2+\mu^2]$ has the same radial form as the timelike-geodesic effective potential. Its outer minimum and inner maximum merge at $r=6M$, the Schwarzschild innermost stable circular orbit, when $(M\mu_c)^2=\ell(\ell+1)/12$. With $\delta=1-(M\mu)^2/(M\mu_c)^2$, we derive exact fold scalings: stationary-point separation $\propto\delta^{1/2}$, barrier--well depth $\propto\delta^{3/2}$, and curvature magnitudes $\propto\delta^{1/2}$. Continued-fraction calculations for the first three axial quasibound branches with $\ell=1,2,3$, cross-checked for selected modes by independent Riccati shooting, show regular frequency continuation across the threshold and approximately linear near-critical behavior. At fixed radius the radial operator is affine in $\delta$, supporting a conditional simple-root analytic interpretation. The main result is the explicit distinction between geometric potential criticality and spectral degeneracy for the branches examined.
\end{abstract}
\fi

\section{Introduction}
\label{sec:introduction}
Massive fields in black-hole spacetimes provide a setting in which the local geometry of an effective scattering potential can be compared with the global analytic properties of its mode spectrum. An exterior maximum or minimum is identified by a radial derivative at a real position, whereas a complex quasibound frequency is selected simultaneously by an ingoing condition at the horizon and decay at infinity. A change in the number of stationary points therefore does not, on its own, establish a spectral degeneracy. The purpose of this work is to quantify that distinction for the decoupled odd-parity sector of a massive vector field in Schwarzschild spacetime.

The interpretation of black-hole resonances as dissipative boundary-value problems and the related numerical methods have a substantial history. Classic treatments introduced the Regge--Wheeler reduction~\cite{ReggeWheeler1957}, the continued-fraction approach~\cite{Leaver1985}, and the frequency-domain decomposition into poles and branch-cut contributions~\cite{Leaver1986}. Reviews by Kokkotas and Schmidt~\cite{KokkotasSchmidt1999}, Nollert~\cite{Nollert1999}, Berti, Cardoso, and Starinets~\cite{BertiCardosoStarinets2009}, and Konoplya and Zhidenko~\cite{KonoplyaZhidenko2011} describe the physical and computational context. These works provide methodological background; our geometric fold is not a new definition of a quasinormal mode.

The fundamental equations and spectrum of Schwarzschild Proca perturbations are established. Rosa and Dolan~\cite{RosaDolan} derived the odd-parity master equation, the coupled even-parity equations, and quasinormal and quasibound spectra using recurrence methods and numerical integration. Earlier work examined massive-vector stability and late-time behavior~\cite{Konoplya2006,KonoplyaZhidenkoMolina}; extensions to rotating spacetimes and alternative spectral techniques are also available~\cite{FrolovEtAl,PercivalDolan,GeMatsumotoZhang}. Our calculation does not claim the discovery of the Proca radial potential, of quasibound states, or of the continued-fraction method. Massive-vector degrees of freedom are also important in rotating-black-hole phenomenology and possible photon-mass constraints~\cite{PaniEtAl2012,BritoCardosoPani2015}; those rotation-induced mechanisms are not part of our static Schwarzschild calculation. Rather, we investigate a particular geometrically defined critical coupling and compare its exact local scaling with the continuation of selected complex modes.

For multipole index $\ell\geq1$, the odd-parity effective potential is $V_{\rm ax}=f(r)[\ell(\ell+1)/r^2+\mu^2]$, with $f=1-2M/r$. Dividing it by $\mu^2$ produces exactly the radial functional form of the classical timelike-geodesic effective potential, with a formal effective specific angular momentum $\mathcal L_{\rm eff}^2=\ell(\ell+1)/\mu^2$. The latter connection is mathematical, not an identification between geodesic trajectories and field solutions. The geodesic interpretation follows the standard discussion of circular Schwarzschild orbits in classical general relativity~\cite{Wald1984}. Its marginally stable circular orbit at $r=6M$ is the well-known Schwarzschild ISCO~\cite{Darwin1959,Chandrasekhar}. Here it identifies the location at which the axial potential maximum and exterior minimum merge as $(M\mu)^2$ reaches $\ell(\ell+1)/12$.

Our first contribution is an exact characterization of this potential fold. If $\delta=1-12(M\mu)^2/[\ell(\ell+1)]$, the stationary-point separation and their curvatures scale as $\delta^{1/2}$, whereas the maximum--minimum depth vanishes as $\delta^{3/2}$. The second contribution is a comparison with the global quasibound spectrum: we continue nine odd-parity modes for $\ell=1,2,3$ across the geometric threshold and test selected roots independently through two-sided Riccati shooting. The complex frequencies remain finite and vary regularly over the branches tested, instead of reproducing the local square-root exponent. Our third contribution is an analytic interpretation. At a fixed radius the differential operator depends affinely on $\delta$; under explicitly stated analyticity and simple-root assumptions, the implicit-function theorem yields a locally analytic spectral branch. We also check that the tracked frequencies stay away from the massive asymptotic branch point.

These claims should be distinguished from recent studies of critical \emph{spectral} masses for scalar fields~\cite{AlvesPonquioMedeiros} and of the two-sheeted massive-field asymptotics~\cite{BaticEtAl2026}. In the present problem the critical coupling is defined first by the stationary-point discriminant of a \emph{real} axial potential. A recent threshold-asymptotics preprint for Proca fields provides a useful complementary context~\cite{Gunara2026}. Related investigations of ringdown spectral stability further emphasize that the appearance of a potential well and the stability of complex-frequency modes need not be equivalent~\cite{WangWu2026}. Our calculations furnish a concrete Schwarzschild spin-one example, while the coupled even-parity sector, possible multiple spectral roots elsewhere, and rigorous global error bounds remain outside the conclusions supported here.

Our use of spherical harmonic sectors is consonant with gauge-invariant master-variable methods for spherically symmetric perturbations~\cite{KodamaIshibashi2003}. This formal comparison does not imply that the coupled Proca even sector is governed by a single scalar master equation.

The paper is organized as follows. Section~\ref{sec2} reviews Maxwell and Proca fields, their constraints and parity decomposition. Sections~\ref{sec10} and~\ref{sec17} develop the radial equations, the potential geometry and the complex quasibound calculation. Section~\ref{sec34} compares scalar and vector field content. The final sections discuss the interpretation, limitations and conclusions; appendices record the angular identities and recurrence used for the numerical spectral calculation.

\section{Field equations, constraints, and harmonic decomposition}
\label{sec2}
\subsection{Conventions and geometric setup}
\label{sec3}

We use the metric signature $(-,+,+,+)$ and geometric units
\begin{eqnarray}
G&=&c=\hbar=1. \label{eq.1}
\end{eqnarray}
Greek indices run over spacetime coordinates and capital Latin indices $A,B,\ldots$ refer to angular coordinates on the unit two-sphere. The antisymmetric field strength is defined by
\begin{eqnarray}
F_{\mu\nu}&=&\nabla_\mu A_\nu-\nabla_\nu A_\mu. \label{eq.2}
\end{eqnarray}
Because the Levi-Civita connection is torsion free, the Christoffel symbols cancel in this antisymmetric combination, so one may equally write
\begin{eqnarray}
F_{\mu\nu}&=&\partial_\mu A_\nu-\partial_\nu A_\mu. \nonumber
\end{eqnarray}

The line element of a generic static, spherically symmetric spacetime will be written as
\begin{eqnarray}
\dd s^2&=&-f(r)\dd t^2+f(r)^{-1}\dd r^2\nonumber\\
&&+r^2\dd\Omega^2, \label{eq.3}\\
\dd\Omega^2&=&\dd\theta^2+\sin^2\theta\,\dd\phi^2. \nonumber
\end{eqnarray}
For Schwarzschild spacetime,
\begin{eqnarray}
f(r)&=&1-\frac{2M}{r}. \label{eq.4}
\end{eqnarray}
The determinant is
\begin{eqnarray}
\sqrt{-g}&=&r^2\sin\theta. \nonumber
\end{eqnarray}

\subsection{Maxwell theory}
\label{sec4}

In vacuum, the Maxwell action is
\begin{eqnarray}
S_{\rm EM}&=&-\frac{1}{4}\int \dd^4x\,\sqrt{-g}\,F_{\mu\nu}F^{\mu\nu}. \label{eq.5}
\end{eqnarray}
To vary the action we use
\begin{eqnarray}
\delta F_{\mu\nu}&=&\nabla_\mu\delta A_\nu-\nabla_\nu\delta A_\mu. \nonumber
\end{eqnarray}
Hence
\begin{eqnarray}
\delta S_{\rm EM}
&=&-\frac{1}{2}\int \dd^4x\sqrt{-g}\,F^{\mu\nu}\delta F_{\mu\nu} \nonumber\\
&=&-\int \dd^4x\sqrt{-g}\,F^{\mu\nu}\nabla_\mu\delta A_\nu. \nonumber
\end{eqnarray}
After an integration by parts and discarding the boundary term,
\begin{eqnarray}
\delta S_{\rm EM}
&=&\int \dd^4x\sqrt{-g}\,(\nabla_\mu F^{\mu\nu})\delta A_\nu. \nonumber
\end{eqnarray}
Since $\delta A_\nu$ is arbitrary, the vacuum Maxwell equation is
\begin{eqnarray}
\boxed{\nabla_\mu F^{\mu\nu}=0}. \label{eq.6}
\end{eqnarray}
The second pair of Maxwell equations follows identically from the definition of $F_{\mu\nu}$:
\begin{eqnarray}
\nabla_{[\alpha}F_{\mu\nu]}&=&0. \label{eq.7}
\end{eqnarray}

A useful feature of the massless theory is that the action is unchanged by
\begin{eqnarray}
A_\mu&\longrightarrow&A_\mu+\nabla_\mu\chi, \label{eq.8}
\end{eqnarray}
for an arbitrary scalar $\chi$. This equivalence relation is part of the physical definition of the massless theory: different potentials related in this way encode the same field strength, and the redundant components must not be counted as independent polarizations.

\subsubsection{Lorenz gauge and the vector wave equation}
\label{sec5}

The Lorenz gauge is
\begin{eqnarray}
\nabla_\mu A^\mu&=&0. \label{eq.9}
\end{eqnarray}
Expanding the Maxwell equation gives
\begin{eqnarray}
\nabla_\mu F^{\mu\nu}
&=&\nabla_\mu\nabla^\mu A^\nu-\nabla_\mu\nabla^\nu A^\mu. \nonumber
\end{eqnarray}
Commuting the derivatives in the second term,
\begin{eqnarray}
\nabla_\mu\nabla^\nu A^\mu
&=&\nabla^\nu(\nabla_\mu A^\mu)+R^{\nu}{}_{\sigma}A^\sigma. \nonumber
\end{eqnarray}
Therefore, in Lorenz gauge,
\begin{eqnarray}
\boxed{\Boxg A^\nu-R^{\nu}{}_{\sigma}A^\sigma=0}. \label{eq.10}
\end{eqnarray}
On a Ricci-flat spacetime such as Schwarzschild, $R_{\mu\nu}=0$, and the potential satisfies a covariant vector wave equation. Even then, the components remain linked by the gauge condition.

\subsection{Proca theory and the dynamical Lorenz constraint}
\label{sec6}

The Proca action is obtained by adding the unique quadratic mass term,
\begin{eqnarray}
S_{\rm P}&=&\int \dd^4x\sqrt{-g}
\times\left[-\frac{1}{4}F_{\mu\nu}F^{\mu\nu}\right.\nonumber\\
&&\qquad\left.-\frac{1}{2}\mu^2 A_\mu A^\mu\right]. \label{eq.11}
\end{eqnarray}
The parameter $\mu$ is the inverse Compton wavelength of the vector excitation in geometric units. Varying $A_\mu$ yields
\begin{eqnarray}
\boxed{\nabla_\mu F^{\mu\nu}-\mu^2A^\nu=0}. \label{eq.12}
\end{eqnarray}
Now take a covariant divergence:
\begin{eqnarray}
\nabla_\nu\nabla_\mu F^{\mu\nu}-\mu^2\nabla_\nu A^\nu&=&0. \nonumber
\end{eqnarray}
The first term vanishes because it is the divergence of an antisymmetric tensor. For $\mu\neq0$ we therefore obtain
\begin{eqnarray}
\boxed{\nabla_\mu A^\mu=0}. \label{eq.13}
\end{eqnarray}
Although Eq.~\eqref{eq.13} has the same algebraic form as the Lorenz condition used in electromagnetism, its status is different. For $\mu\neq0$ it is obtained by taking the divergence of the equations of motion; no gauge transformation is available to impose or undo it. It is therefore a dynamical constraint rather than a gauge convention.

Using Eq.~\eqref{eq.13} in Eq.~\eqref{eq.12}, one finds
\begin{eqnarray}
\Boxg A^\nu-R^{\nu}{}_{\sigma}A^\sigma-\mu^2A^\nu&=&0. \label{eq.14}
\end{eqnarray}
In Ricci-flat spacetime this becomes
\begin{eqnarray}
(\Boxg-\mu^2)A^\nu&=&0, \nonumber
\end{eqnarray}
subject to the non-gauge Lorenz constraint.

\subsection{Degrees of freedom and the flat-spacetime limit}
\label{sec7}

The polarization count is the first place where the two theories part company in an observable way. A Maxwell potential begins with four components, but the constraint structure together with gauge equivalence leaves two propagating transverse states. A Proca field also begins with four components, yet there is no gauge orbit to quotient out. Equation~\eqref{eq.13} removes one independent component and leaves three physical modes, which may be viewed as two transverse polarizations plus one longitudinal polarization.

The difference can be summarized as
\begin{eqnarray}
N_{\rm Maxwell}&=&2,\qquad N_{\rm Proca}=3. \label{eq.15}
\end{eqnarray}
This counting reappears directly in the parity decomposition: the Maxwell even sector contains a gauge redundancy, whereas the corresponding Proca amplitude contributes to a physical longitudinal mode.

\paragraph{Plane-wave limit.}

The flat-space plane-wave limit provides a useful diagnostic before curvature and angular momentum are introduced. Let
\begin{eqnarray}
A^\mu(x)&=&\epsilon^\mu e^{-\ii k_\alpha x^\alpha}. \nonumber
\end{eqnarray}
For Proca theory, Eq.~\eqref{eq.14} gives
\begin{eqnarray}
k_\mu k^\mu+\mu^2&=&0, \label{eq.16}
\end{eqnarray}
and the constraint gives
\begin{eqnarray}
k_\mu\epsilon^\mu&=&0. \label{eq.17}
\end{eqnarray}
Writing $k^\mu=(\omega,\mathbf{k})$ in flat spacetime,
\begin{eqnarray}
\boxed{\omega^2=\mathbf{k}^2+\mu^2}. \label{eq.18}
\end{eqnarray}
Maxwell theory is recovered at the level of the dispersion relation by setting $\mu=0$, so $\omega=|\mathbf{k}|$. The important caveat is that the number of physical polarizations changes discontinuously because gauge invariance is restored at exactly $\mu=0$.

\begin{figure}[t]
\centering
\includegraphics[width=0.98\columnwidth]{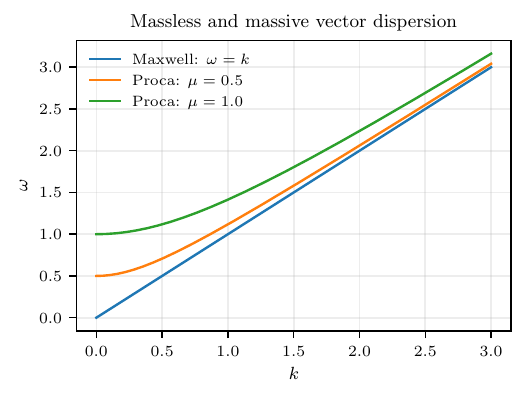}
\caption{Flat-spacetime dispersion relations. The Proca mass opens a gap at zero momentum, whereas the Maxwell field remains gapless.}
\label{fig:dispersion}
\end{figure}

\subsection{Vector spherical harmonics}
\label{sec8}

Ordinary spherical harmonics are sufficient for a scalar, but a vector potential also has components tangent to the two-sphere. The angular basis must therefore contain covectors/vectors on $S^2$, not only scalar functions. Let $D_A$ be the covariant derivative compatible with the unit-sphere metric $\Omega_{AB}$, and let $\epsilon_{AB}$ be its Levi-Civita tensor. Starting from $Y_{\ell m}$ we define two angular vectors,
\begin{eqnarray}
Y_A^{\ell m}&=&D_A Y_{\ell m}, \label{eq.19}\\
X_A^{\ell m}&=&-\epsilon_A{}^{B}D_B Y_{\ell m}. \label{eq.20}
\end{eqnarray}
The first is polar (even parity) and the second is axial (odd parity). The scalar harmonic obeys
\begin{eqnarray}
D^AD_A Y_{\ell m}&=&-\ell(\ell+1)Y_{\ell m}. \label{eq.21}
\end{eqnarray}
From this definition,
\begin{eqnarray}
D^A X_A^{\ell m}&=&0, \nonumber\\
D^A Y_A^{\ell m}&=&-\ell(\ell+1)Y_{\ell m}. \label{eq.22}
\end{eqnarray}
Under parity $(\theta,\phi)\rightarrow(\pi-\theta,\phi+\pi)$,
\begin{eqnarray}
Y_A^{\ell m}&\rightarrow&(-1)^\ell Y_A^{\ell m},\nonumber\\
X_A^{\ell m}&\rightarrow&(-1)^{\ell+1}X_A^{\ell m}. \label{eq.23}
\end{eqnarray}
The parity split is powerful because a spherically symmetric background does not mix these two sectors.

\subsection{Parity decomposition of the vector potential}
\label{sec9}

Separate spacetime indices into $a,b\in\{t,r\}$ and angular indices $A,B\in\{\theta,\phi\}$. A general mode of the vector potential may be written as
\begin{eqnarray}
A_a&=&a_a(t,r)Y_{\ell m}, \nonumber\\
A_A&=&a_{\rm e}(t,r)Y_A^{\ell m}+a_{\rm o}(t,r)X_A^{\ell m}. \label{eq.24}
\end{eqnarray}
The functions $a_t,a_r,a_{\rm e}$ belong to the even sector, and $a_{\rm o}$ belongs to the odd sector.

For Maxwell theory a gauge transformation generated by
\begin{eqnarray}
\chi&=&\chi_{\ell m}(t,r)Y_{\ell m} \nonumber
\end{eqnarray}
acts as
\begin{eqnarray}
a_t&\rightarrow&a_t+\partial_t\chi_{\ell m},\nonumber\\
a_r&\rightarrow&a_r+\partial_r\chi_{\ell m},\nonumber\\
a_{\rm e}&\rightarrow&a_{\rm e}+\chi_{\ell m},\nonumber\\
a_{\rm o}&\rightarrow&a_{\rm o}. \label{eq.25}
\end{eqnarray}
Thus the odd amplitude is automatically gauge invariant. The axial amplitude is therefore gauge invariant already at the level of the harmonic decomposition.

\section{Radial dynamics in Schwarzschild spacetime}
\label{sec10}
\subsection{Maxwell odd-parity sector}
\label{sec11}

For an odd-parity electromagnetic mode we take
\begin{eqnarray}
A_t&=&0,\qquad A_r=0,\nonumber\\
A_A&=&a_{\rm o}(t,r)X_A^{\ell m}. \label{eq.26}
\end{eqnarray}
The nonzero field-strength components are
\begin{eqnarray}
F_{tA}&=&\partial_t a_{\rm o}\,X_A^{\ell m}, \nonumber\\
F_{rA}&=&\partial_r a_{\rm o}\,X_A^{\ell m}, \nonumber
\end{eqnarray}
and the purely angular part is proportional to the curl of $X_A^{\ell m}$. Substituting the decomposition into Eq.~\eqref{eq.6}, using the angular identities, and projecting onto $X_A^{\ell m}$ gives
\begin{eqnarray}
-\frac{1}{f}\frac{\partial^2 a_{\rm o}}{\partial t^2}
+\frac{\partial}{\partial r}\left(f\frac{\partial a_{\rm o}}{\partial r}\right)&=&\nonumber\\
&&\frac{\ell(\ell+1)}{r^2}a_{\rm o}. \label{eq.27}
\end{eqnarray}
Introduce the tortoise coordinate $r_*$ by
\begin{eqnarray}
\frac{\dd r_*}{\dd r}&=&\frac{1}{f(r)}, \label{eq.28}
\end{eqnarray}
so that
\begin{eqnarray}
\frac{\partial}{\partial r_*}&=&f\frac{\partial}{\partial r}. \nonumber
\end{eqnarray}
Equation~\eqref{eq.27} becomes
\begin{eqnarray}
\left[-\frac{\partial^2}{\partial t^2}
+\frac{\partial^2}{\partial r_*^2}\right]a_{\rm o}&=&\nonumber\\
&&V_{\rm EM}(r)\,a_{\rm o}. \label{eq.29}
\end{eqnarray}
with
\begin{eqnarray}
\boxed{V_{\rm EM}(r)=f(r)\frac{\ell(\ell+1)}{r^2}}. \label{eq.30}
\end{eqnarray}
For harmonic time dependence
\begin{eqnarray}
a_{\rm o}(t,r)&=&e^{-\ii\omega t}\psi_{\rm o}(r), \nonumber
\end{eqnarray}
we obtain
\begin{eqnarray}
\frac{\dd^2\psi_{\rm o}}{\dd r_*^2}+
\left[\omega^2-V_{\rm EM}(r)\right]\psi_{\rm o}&=&0. \label{eq.31}
\end{eqnarray}
This is the spin-1 Regge--Wheeler equation, in agreement with the standard Schwarzschild electromagnetic perturbation treatment \cite{Chandrasekhar}.

\subsection{Maxwell even-parity sector and parity equivalence}
\label{sec12}

The even sector contains $a_t$, $a_r$, and $a_{\rm e}$. At first sight it looks more complicated because gauge freedom acts directly on all three amplitudes. A convenient gauge-invariant combination is the radial electric field amplitude,
\begin{eqnarray}
\mathcal{E}_{\ell m}(t,r)&=&r^2\left(\partial_t a_r-\partial_r a_t\right). \label{eq.32}
\end{eqnarray}
After inserting the even decomposition into Maxwell's equations, the angular equation acts as a constraint and the remaining equations may be combined into a single master equation. With an appropriate normalization $\psi_{\rm e}\propto\mathcal{E}_{\ell m}$, one finds
\begin{eqnarray}
\frac{\dd^2\psi_{\rm e}}{\dd r_*^2}+
\left[\omega^2-V_{\rm EM}(r)\right]\psi_{\rm e}&=&0. \label{eq.33}
\end{eqnarray}
The even and odd electromagnetic sectors therefore share the same effective potential. Their equality is not an accident; it reflects electromagnetic duality and the absence of an independent longitudinal photon polarization.

\paragraph{Electromagnetic monopole.}

The vector harmonic $X_A^{00}$ vanishes because $Y_{00}$ is constant. The radiative Maxwell sectors therefore begin at $\ell=1$. The $\ell=0$ solution is the static Coulomb field. In Schwarzschild coordinates it may be written as
\begin{eqnarray}
A_t&=&\frac{Q}{r}+{\rm const.}, \label{eq.34}
\end{eqnarray}
with
\begin{eqnarray}
F_{tr}&=&\frac{Q}{r^2}. \nonumber
\end{eqnarray}
Thus the electromagnetic monopole carries charge information rather than a radiative degree of freedom.

\subsection{Proca harmonic amplitudes}
\label{sec13}

The same vector spherical harmonics can be used for the massive field, but there is now no gauge transformation to remove an even-parity amplitude. Following the conventional normalization used in the Schwarzschild Proca literature, one may expand
\begin{eqnarray}
A_\mu&=&\frac{u_{(1)}}{r}Y_{\ell m}\,\delta^t_\mu
+\frac{u_{(2)}}{rf}Y_{\ell m}\,\delta^r_\mu\nonumber\\
&&+\frac{u_{(3)}}{\ell(\ell+1)}Y_A^{\ell m}\,\delta^A_\mu\nonumber\\
&&+\frac{u_{(4)}}{\ell(\ell+1)}X_A^{\ell m}\,\delta^A_\mu. \label{eq.35}
\end{eqnarray}
Here $u_{(1)},u_{(2)},u_{(3)}$ are even-parity amplitudes and $u_{(4)}$ is odd parity. Different papers use slightly different normalizations, but the physical content is the same; the present convention is aligned with the Schwarzschild Proca analysis of Rosa and Dolan \cite{RosaDolan}.

Define the differential operator
\begin{eqnarray}
\mathcal{D}_s&=&\frac{\partial^2}{\partial r_*^2}
-\frac{\partial^2}{\partial t^2}\nonumber\\
&&-f\left[\frac{\ell(\ell+1)}{r^2}+\mu^2
+\frac{1-s^2}{r}\frac{\dd f}{\dd r}\right]. \label{eq.36}
\end{eqnarray}
For $s=1$ the curvature derivative term drops out. This compact notation will make the relation to Maxwell theory especially clear.

\subsection{Proca odd-parity sector}
\label{sec14}

The axial amplitude $u_{(4)}$ decouples completely. Its equation is
\begin{eqnarray}
\boxed{\mathcal{D}_1u_{(4)}=0}. \label{eq.37}
\end{eqnarray}
Using Eq.~\eqref{eq.36},
\begin{eqnarray}
\left[-\frac{\partial^2}{\partial t^2}
+\frac{\partial^2}{\partial r_*^2}\right]u_{(4)}&=&
f\left(\frac{\ell(\ell+1)}{r^2}+\mu^2\right)u_{(4)}.\nonumber\\
\label{eq.38}
\end{eqnarray}
For $u_{(4)}=e^{-\ii\omega t}\psi_{(-)}(r)$,
\begin{eqnarray}
\frac{\dd^2\psi_{(-)}}{\dd r_*^2}+
\left[\omega^2-V_{\rm P}^{(-)}(r)\right]\psi_{(-)}&=&0, \label{eq.39}
\end{eqnarray}
where
\begin{eqnarray}
\boxed{V_{\rm P}^{(-)}(r)=f(r)\left[\frac{\ell(\ell+1)}{r^2}+\mu^2\right]}. \label{eq.40}
\end{eqnarray}
Equation~\eqref{eq.40} is the cleanest bridge between Maxwell and Proca theory. The angular barrier is unchanged; the vector mass adds a new positive term inside the square brackets.

\begin{figure}[t]
\centering
\includegraphics[width=0.98\columnwidth]{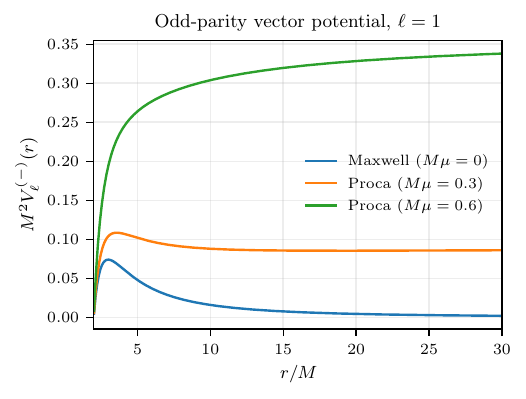}
\caption{Odd-parity effective potential for $\ell=1$ and $M=1$. The Maxwell curve corresponds to $\mu=0$. A nonzero Proca mass changes the asymptotic value of the potential from zero to $\mu^2$.}
\label{fig:potentialmass}
\end{figure}

\subsection{Proca even-parity sector}
\label{sec15}

In the electric-parity sector the longitudinal Proca degree of freedom is explicit. The amplitudes $u_{(1)},u_{(2)},u_{(3)}$ obey the coupled system \cite{RosaDolan}
\begin{eqnarray}
\mathcal{D}_1u_{(1)}
+\frac{\dd f}{\dd r}\left(\dot u_{(2)}-u_{(1)}'\right)&=&0, \label{eq.41}\\
\mathcal{D}_1u_{(2)}
+\frac{\dd f}{\dd r}\left(\dot u_{(1)}-u_{(2)}'\right)&=&\nonumber\\
&&\frac{2f^2}{r^2}\left(u_{(2)}\right.\nonumber\\
&&\left.\qquad-u_{(3)}\right), \label{eq.42}\\
\mathcal{D}_1u_{(3)}
+\frac{2f\ell(\ell+1)}{r^2}u_{(2)}&=&0. \label{eq.43}
\end{eqnarray}
A dot denotes $\partial_t$ and a prime denotes $\partial_{r_*}$. The Lorenz constraint is
\begin{eqnarray}
-\dot u_{(1)}+u_{(2)}'+\frac{f}{r}\left(u_{(2)}-u_{(3)}\right)&=&0. \label{eq.44}
\end{eqnarray}
Using Eq.~\eqref{eq.44} to eliminate $\dot u_{(1)}$ from Eq.~\eqref{eq.42}, one obtains the two-field system
\begin{eqnarray}
\mathcal{D}_1u_{(2)}
+\frac{f}{r}\left(\frac{\dd f}{\dd r}-\frac{2f}{r}\right)
\left(u_{(2)}-u_{(3)}\right)&=&0, \label{eq.45}\\
\mathcal{D}_1u_{(3)}
+\frac{2f\ell(\ell+1)}{r^2}u_{(2)}&=&0. \label{eq.46}
\end{eqnarray}
The field therefore contains two even-parity physical polarizations plus the one odd-parity polarization in Eq.~\eqref{eq.37}. The count agrees with Eq.~\eqref{eq.15}.

\subsection{Constraint structure in the massless limit}
\label{sec16}

Because the Proca equations contain Maxwell's equations when the explicit mass term is removed, it is tempting to regard the limit $\mu\rightarrow0$ as a purely algebraic substitution. The differential equations do approach massless equations, but the constraint structure also changes. When $\mu\neq0$, Eq.~\eqref{eq.13} is enforced by the field equations. At $\mu=0$, the same condition may instead be imposed as a gauge choice, and a new gauge redundancy appears.

The symmetry content changes discontinuously at the endpoint of the limit. The longitudinal massive mode is not an additional photon polarization once gauge equivalence is restored. The limit
\begin{eqnarray}
\mu&\rightarrow&0 \label{eq.47}
\end{eqnarray}
is dynamically smooth in the odd sector, where Eq.~\eqref{eq.40} becomes Eq.~\eqref{eq.30}, but it is conceptually nontrivial in the even sector because the number of physical degrees of freedom changes from three to two.

\paragraph{Proca monopole.}

For $\ell=0$, the angular vector harmonics vanish. Consequently $u_{(3)}$ and $u_{(4)}$ are absent, but the massive theory still possesses a dynamical monopole associated with the longitudinal polarization. The remaining radial variable can be arranged to satisfy
\begin{eqnarray}
\frac{\partial^2u_{(2)}}{\partial r_*^2}
-\frac{\partial^2u_{(2)}}{\partial t^2}&=&\nonumber\\
&&f\left[\mu^2+\frac{2}{r^2}\left(1-\frac{3M}{r}\right)\right]\nonumber\\
&&\quad\times u_{(2)}.\label{eq.48}
\end{eqnarray}
This mode has no radiative Maxwell counterpart. In the massless theory, the monopole reduces to the static Coulomb sector plus a pure-gauge contribution. In the massive theory the longitudinal monopole is physical.

\section{Asymptotics, effective potentials, and spectra}
\label{sec17}
\subsection{Near-horizon behavior}
\label{sec18}

At the event horizon $r\rightarrow2M$, one has
\begin{eqnarray}
f(r)&\rightarrow&0. \nonumber
\end{eqnarray}
Hence both the Maxwell potential and the Proca odd-parity potential vanish:
\begin{eqnarray}
V_{\rm EM}&\rightarrow&0,\qquad V_{\rm P}^{(-)}\rightarrow0. \label{eq.49}
\end{eqnarray}
The master equations therefore reduce locally to
\begin{eqnarray}
\frac{\dd^2\psi}{\dd r_*^2}+\omega^2\psi&\simeq&0. \nonumber
\end{eqnarray}
The two elementary behaviors are
\begin{eqnarray}
\psi&\sim&e^{\pm\ii\omega r_*}. \label{eq.50}
\end{eqnarray}
With the convention $e^{-\ii\omega t}$, the future-horizon physical solution is the ingoing wave $e^{-\ii\omega(t+r_*)}$.

\subsection{Asymptotic infinity}
\label{sec19}

For Maxwell theory,
\begin{eqnarray}
V_{\rm EM}(r)&\longrightarrow&0,\qquad r\rightarrow\infty. \nonumber
\end{eqnarray}
Thus the far-zone wavenumber is simply $k=\omega$. For the odd Proca mode,
\begin{eqnarray}
V_{\rm P}^{(-)}(r)&\longrightarrow&\mu^2, \label{eq.51}
\end{eqnarray}
so the radial equation becomes
\begin{eqnarray}
\frac{\dd^2\psi_{(-)}}{\dd r_*^2}+k_\infty^2\psi_{(-)}&\simeq&0, \nonumber\\
k_\infty^2&=&\omega^2-\mu^2. \label{eq.52}
\end{eqnarray}
If $\omega^2>\mu^2$, the far solution oscillates. If $\omega^2<\mu^2$, then $k_\infty$ is imaginary and the field can decay exponentially at infinity. This asymptotic mass gap permits exponentially decaying boundary conditions and hence quasibound states with no Maxwell analogue.

\subsection{Multipole dependence and spin structure}
\label{sec20}

The same two-sphere eigenvalue controls the angular barrier in both vector theories; the distinction is carried by the remaining radial terms and by the number of independent amplitudes. For the odd Proca sector,
\begin{eqnarray}
V_{\ell}^{(-)}(r)&=&f(r)\left[\frac{\ell(\ell+1)}{r^2}+\mu^2\right]. \nonumber
\end{eqnarray}
Increasing $\ell$ raises the barrier near the black hole while leaving the asymptotic plateau $\mu^2$ unchanged.

\begin{figure}[t]
\centering
\includegraphics[width=0.98\columnwidth]{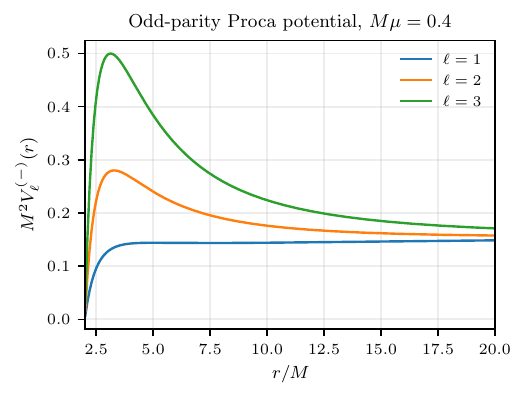}
\caption{Odd-parity Proca potential for several multipoles at fixed $\mu M=0.4$. The angular barrier grows rapidly with $\ell$, whereas all curves approach the same mass plateau at large radius.}
\label{fig:multipoles}
\end{figure}

\paragraph{Massless spin potential.}

For massless test fields of spin $s$ in Schwarzschild spacetime, a useful family of Regge--Wheeler-type potentials is
\begin{eqnarray}
V_s(r)&=&f(r)\left[\frac{\ell(\ell+1)}{r^2}+(1-s^2)\frac{2M}{r^3}\right]. \label{eq.53}
\end{eqnarray}
For a scalar field, $s=0$, and the curvature term $2M/r^3$ survives. For the electromagnetic field, $s=1$, so the curvature term cancels:
\begin{eqnarray}
V_{s=1}(r)&=&f(r)\frac{\ell(\ell+1)}{r^2}. \nonumber
\end{eqnarray}
This identity is a useful consistency check on the electromagnetic equation. It should not, however, be promoted to a universal recipe for a massive vector: the odd Proca mode admits a simple potential, while the electric sector retains the coupling associated with the longitudinal degree of freedom.

\subsection{Analytic structure of the radial barrier}
\label{sec21}
\paragraph{Maxwell barrier.}

The Maxwell potential is
\begin{eqnarray}
V_{\rm EM}(r)&=&\ell(\ell+1)\left(\frac{1}{r^2}-\frac{2M}{r^3}\right). \nonumber
\end{eqnarray}
The extremum satisfies
\begin{eqnarray}
\frac{\dd V_{\rm EM}}{\dd r}&=&\ell(\ell+1)
\left(-\frac{2}{r^3}+\frac{6M}{r^4}\right)=0. \nonumber
\end{eqnarray}
Multiplying by $r^4/[2\ell(\ell+1)]$ gives
\begin{eqnarray}
-r+3M&=&0. \nonumber
\end{eqnarray}
Therefore
\begin{eqnarray}
\boxed{r_{\rm peak}^{\rm EM}=3M}. \label{eq.54}
\end{eqnarray}
The Maxwell barrier peaks at the Schwarzschild photon sphere. In Proca theory the term $f\mu^2$ changes the radial slope and shifts the stationary points.

\paragraph{Axial Proca extrema.}

For the odd mode, define
\begin{eqnarray}
L&=&\ell(\ell+1). \nonumber
\end{eqnarray}
Then
\begin{eqnarray}
V(r)&=&\left(1-\frac{2M}{r}\right)\left(\frac{L}{r^2}+\mu^2\right). \nonumber
\end{eqnarray}
Differentiate:
\begin{eqnarray}
\frac{\dd V}{\dd r}
&=&\frac{2M}{r^2}\left(\frac{L}{r^2}+\mu^2\right)
+\left(1-\frac{2M}{r}\right)\left(-\frac{2L}{r^3}\right). \nonumber
\end{eqnarray}
Multiplying by $r^4/2$ gives
\begin{eqnarray}
M\left(L+\mu^2r^2\right)-Lr+2ML&=&0. \nonumber
\end{eqnarray}
Hence the stationary radius satisfies
\begin{eqnarray}
M\mu^2r^2-Lr+3ML&=&0. \label{eq.55}
\end{eqnarray}
For $\mu\rightarrow0$ the finite root approaches $3M$, as required. Introducing
\begin{eqnarray}
x&=&\frac{r}{M},\qquad \alpha=M\mu, \qquad L=\ell(\ell+1), \nonumber
\end{eqnarray}
Eq.~\eqref{eq.55} becomes
\begin{eqnarray}
\alpha^2x^2-Lx+3L&=&0. \label{eq.56}
\end{eqnarray}
The two stationary radii are therefore
\begin{eqnarray}
x_{\pm}&=&\frac{L\pm\sqrt{L^2-12L\alpha^2}}{2\alpha^2}. \label{eq.57}
\end{eqnarray}
Real stationary points exist only when
\begin{eqnarray}
\alpha^2&\leq&\frac{L}{12}. \label{eq.58}
\end{eqnarray}
For subcritical coupling, $x_{-}$ is the inner maximum and $x_{+}$ is an outer minimum. Their limiting behavior is
\begin{eqnarray}
x_-&=&3+\frac{9\alpha^2}{L}+\mathcal{O}(\alpha^4),\nonumber\\
x_+&=&\frac{L}{\alpha^2}-3+\mathcal{O}(\alpha^2). \label{eq.59}
\end{eqnarray}
At the critical coupling $\alpha_c=\sqrt{L/12}$ the two stationary points merge at $x=6$. For $\alpha>\alpha_c$ the odd-parity potential rises monotonically from the horizon toward its asymptotic value $\mu^2$. Using the stationarity condition, the potential at either extremum can be written compactly as
\begin{eqnarray}
M^2V_{\rm ext}&=&L\frac{(x_{\rm ext}-2)^2}{x_{\rm ext}^3}. \label{eq.60}
\end{eqnarray}
Thus the Maxwell photon-sphere barrier is recovered continuously on the $x_-$ branch, while the Proca mass generates a second, large-radius stationary point associated with the approach to the massive asymptotic plateau.

\begin{figure}[t]
\centering
\includegraphics[width=0.98\columnwidth]{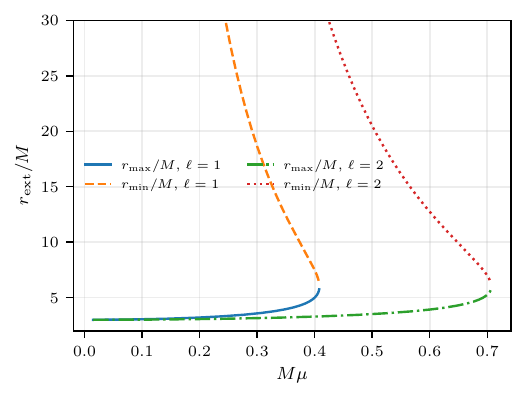}
\caption{Stationary radii of the odd-parity Proca potential as functions of $M\mu$ for $\ell=1$ and $\ell=2$. The lower branch is the barrier maximum continuously connected to $r=3M$ in the Maxwell limit; the upper branch is an outer minimum. The two branches merge at $M\mu=\sqrt{\ell(\ell+1)/12}$ and $r=6M$.}
\label{fig:axial-extrema}
\end{figure}

\subsection{Geodesic correspondence and the meaning of $r=6M$}
\label{sec22}

The appearance of $6M$ can be understood without treating it as an independent new black-hole length scale. Dividing the odd-parity Proca potential by $\mu^2$ gives
\begin{eqnarray}
\frac{V_{\rm P}^{(-)}}{\mu^2}
&=&f(r)\left[1+\frac{\mathcal{L}_{\rm eff}^2}{r^2}\right],
\qquad
\mathcal{L}_{\rm eff}^2\equiv\frac{L}{\mu^2}. \label{eq.61}
\end{eqnarray}
The expression on the right-hand side is precisely the standard effective potential entering the radial equation for timelike geodesics in Schwarzschild spacetime, with $\mathcal{L}_{\rm eff}$ playing the role of the specific angular momentum \cite{Darwin1959,Chandrasekhar}. This is a formal correspondence between radial potentials; it does not identify the Proca wave equation with the geodesic equation.

\begin{figure}[t]
\centering
\includegraphics[width=0.96\columnwidth]{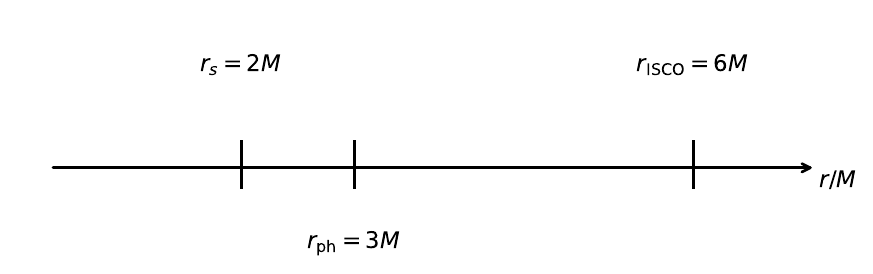}
\caption{Schematic Schwarzschild radial-coordinate markers (not to scale): horizon $r_s=2M$, photon circular orbit $r_{\rm ph}=3M$, and timelike ISCO $r_{\rm ISCO}=6M$. The odd-parity Proca potential extrema coalesce at $r=6M$ when $(M\mu_c)^2=\ell(\ell+1)/12$. The equality of radii follows from the formal effective-potential correspondence; no accretion-disk model is assumed.}
\label{fig:isco-schematic}
\end{figure}

For Schwarzschild timelike geodesics, the two circular-orbit extrema merge at the innermost stable circular orbit, where
\begin{eqnarray}
r_{\rm ISCO}&=&6M,
\qquad
\mathcal{L}_{\rm ISCO}^2=12M^2. \label{eq.62}
\end{eqnarray}
Using Eq.~\eqref{eq.61}, the ISCO angular-momentum condition becomes
\begin{eqnarray}
\frac{L}{\mu_c^2}&=&12M^2,
\end{eqnarray}
which is equivalent to
\begin{eqnarray}
(M\mu_c)^2&=&\frac{L}{12}. \label{eq.63}
\end{eqnarray}
Figure~\ref{fig:isco-schematic} summarizes the established hierarchy of Schwarzschild radii: the event horizon at $r_s=2M$, the null circular orbit at $r=3M=1.5r_s$, and the timelike ISCO at $r=6M=3r_s$. The schematic is a radial-coordinate diagram, not an embedding diagram, a spatially scaled image, or an accretion-disk model. The coincidence of the axial potential fold with the ISCO radius expresses the effective-potential mapping, not an identification of Proca wave propagation with particle geodesics.

Thus the critical coupling obtained from the discriminant of the Proca stationary-point equation is the direct image of the Schwarzschild ISCO condition under the potential map. This observation also clarifies the limiting branches: the inner Proca extremum tends to the null circular orbit at $3M$ as $\mu\to0$, whereas the finite-coupling fold occurs at the timelike marginal-stability radius $6M$.

\subsection{Critical scaling of the axial barrier--well pair}
\label{sec23}

The discriminant in Eq.~\eqref{eq.57} permits a useful exact reparametrization of the two stationary radii. Define
\begin{eqnarray}
\delta&=&1-\frac{12\alpha^2}{L},\qquad s=\sqrt{\delta}, \qquad 0\leq s\leq1. \label{eq.64}
\end{eqnarray}
Then the two roots can be written without a cancellation-prone quadratic expression,
\begin{eqnarray}
x_-&=&\frac{6}{1+s},\qquad x_+=\frac{6}{1-s}. \label{eq.65}
\end{eqnarray}
The separation between the maximum and minimum is therefore
\begin{eqnarray}
\Delta x&\equiv&x_+-x_- = \frac{12s}{1-s^2}, \label{eq.66}
\end{eqnarray}
so that $\Delta x\sim12\delta^{1/2}$ as the critical coupling is approached from below.

A stronger statement follows by evaluating Eq.~\eqref{eq.60} at the two branches. The barrier--well depth is exactly
\begin{eqnarray}
M^2\left(V_{\max}-V_{\min}\right)&=&\frac{L}{27}s^3\nonumber\\
&=&\frac{L}{27}\left(1-\frac{12\alpha^2}{L}\right)^{3/2}. \label{eq.67}
\end{eqnarray}
Thus the height difference vanishes with exponent $3/2$. The second derivative with respect to the dimensionless tortoise coordinate $x_*=r_*/M$ is, at the two extrema,
\begin{eqnarray}
\left.\frac{\dd^2(M^2V)}{\dd x_*^2}\right|_{x_-}
&=&-\frac{Ls(s-2)^2(s+1)^4}{5832},\nonumber\\
\left.\frac{\dd^2(M^2V)}{\dd x_*^2}\right|_{x_+}
&=&\frac{Ls(s-1)^4(s+2)^2}{5832}. \label{eq.68}
\end{eqnarray}
Near criticality these reduce to $\mp L\sqrt{\delta}/1458+\mathcal{O}(\delta)$, so both the barrier curvature and the well curvature vanish as $\delta^{1/2}$. Equations~\eqref{eq.66}--\eqref{eq.68} show that the coalescence at $r=6M$ is controlled by a single square-root parameter, while the potential-depth scale is parametrically smaller, of order $\delta^{3/2}$.

\begin{figure}[t]
\centering
\includegraphics[width=0.98\columnwidth]{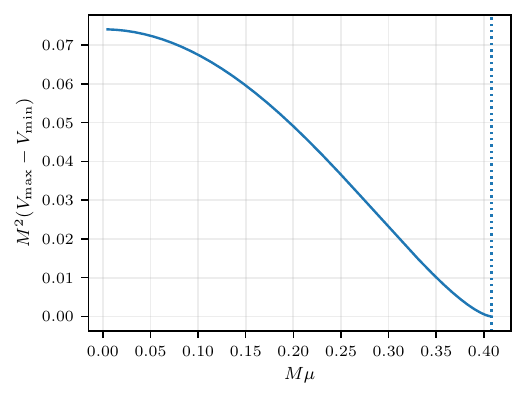}
\caption{Exact barrier--well depth of the odd-parity Proca potential for $\ell=1$. The depth vanishes as $[1-6(M\mu)^2]^{3/2}$ at the critical coupling $M\mu=1/\sqrt{6}$.}
\label{fig:critical-depth}
\end{figure}

\begingroup
\subsection{Analyticity of the potential operator versus nonanalytic stationary radii}
\label{sec24}
The critical exponent $1/2$ describing the stationary radii does not reflect a singular dependence of the wave operator on the Proca mass. This can be seen directly, without solving the wave equation. Set $x=r/M$, $L=\ell(\ell+1)$, and $\delta=1-12(M\mu)^2/L$. In these variables the dimensionless axial potential is
\begin{eqnarray}
U(x,\delta)&\equiv&M^2V(x,\delta)\nonumber\\
&=&\left(1-\frac{2}{x}\right)\left[\frac{L}{x^2}+\frac{L}{12}(1-\delta)\right]\nonumber\\
&=&\underbrace{\left(1-\frac{2}{x}\right)\left(\frac{L}{x^2}+\frac{L}{12}\right)}_{U_c(x)}\nonumber\\
&&-\frac{L}{12}\left(1-\frac{2}{x}\right)\delta.
\label{eq.69}
\end{eqnarray}
Differentiating at fixed $x$, with no approximation, gives
\begin{eqnarray}
\frac{\partial U}{\partial\delta}&=&-\frac{L}{12}\left(1-\frac{2}{x}\right),\nonumber\\
\frac{\partial^2U}{\partial\delta^2}&=&0.\label{eq.70}
\end{eqnarray}
The tortoise coordinate $x_*$ is independent of the field mass. Hence the differential operator $\mathcal{L}=\partial_{x_*}^2+\Omega^2-U(x,\delta)$, with $\Omega=M\omega$, obeys
\begin{eqnarray}
\mathcal{L}(\Omega,\delta)&=&\mathcal{L}(\Omega,0)+\delta\,\frac{L}{12}\left(1-\frac{2}{x}\right),\nonumber\\
\frac{\partial^2\mathcal{L}}{\partial\delta^2}&=&0.
\label{eq.71}
\end{eqnarray}
The roots of $\partial_xU=0$, however, involve the vanishing discriminant $\delta$ and therefore separate as $\sqrt\delta$. Consequently the fold is a ramification of the map from coupling to stationary-point locations, not a singular coefficient of the original wave equation. This statement alone does not prove regularity of the boundary-value spectrum, because the asymptotic boundary conditions also depend on frequency and field mass.

To make the latter restriction explicit, write the quasibound condition as $\mathcal F(\Omega,\delta)=0$ on a fixed decaying asymptotic sheet, assume its local analyticity, and require $\mathcal F_\Omega(\Omega_c,0)\ne0$. Differentiating once along a root yields
\begin{eqnarray}
0&=&\frac{\dd}{\dd\delta}\mathcal F[\Omega(\delta),\delta]\nonumber\\
&=&\mathcal F_\Omega\frac{\dd\Omega}{\dd\delta}+\mathcal F_\delta,
\end{eqnarray}
\begin{eqnarray}
\left.\frac{\dd\Omega}{\dd\delta}\right|_0
&=&-\left.\frac{\mathcal F_\delta}{\mathcal F_\Omega}\right|_{(\Omega_c,0)}.
\label{eq.72}
\end{eqnarray}
Differentiating a second time gives, without omitting the mixed derivative,
\begin{eqnarray}
0&=&\mathcal F_{\Omega\Omega}(\Omega')^2+2\mathcal F_{\Omega\delta}\Omega'\nonumber\\
&&+\mathcal F_{\Omega}\Omega''+\mathcal F_{\delta\delta},
\end{eqnarray}
\begin{eqnarray}
\left.\Omega''\right|_0
&=&-\left.\frac{\mathcal F_{\Omega\Omega}(\Omega')^2}{\mathcal F_\Omega}\right|_{(\Omega_c,0)}\nonumber\\
&&-\left.\frac{2\mathcal F_{\Omega\delta}\Omega'}{\mathcal F_\Omega}\right|_{(\Omega_c,0)}\nonumber\\
&&-\left.\frac{\mathcal F_{\delta\delta}}{\mathcal F_\Omega}\right|_{(\Omega_c,0)}.
\label{eq.73}
\end{eqnarray}
Thus $\Omega(\delta)=\Omega_c+\Omega'_c\delta+\frac12\Omega''_c\delta^2+\cdots$ for an isolated simple root under the stated analyticity assumptions. The numerical fits below are evidence for this prediction on the branches tested, not a theorem excluding multiple roots or threshold collisions for all Proca modes.

\begin{figure}[t]
\centering
\includegraphics[width=0.98\columnwidth]{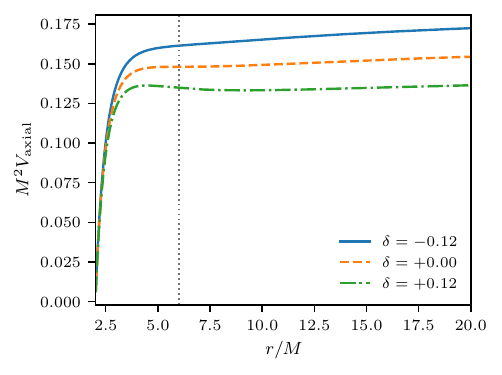}
\caption{Exact, fixed-radius variation of the dimensionless axial potential relative to its critical value for $\ell=1$. The affine dependence on $\delta$ remains smooth even though the locations of the extrema coalesce with square-root scaling. The curves are generated from Eq.~\eqref{eq.69}, not fitted to spectral data.}
\label{fig:affine-operator}
\end{figure}

\endgroup
\subsection{Odd-parity quasibound spectrum near the potential coalescence}
\label{sec25}

To determine how the spectral problem responds to the potential coalescence, we solve the decoupled odd-parity equation with the standard scalar continued-fraction recurrence used for Schwarzschild Proca bound states \cite{RosaDolan}. Existing studies already establish axial quasibound branches across a broad range of $M\mu$ \cite{RosaDolan,GeMatsumotoZhang}; our numerical calculation is therefore used here as a probe of the fold identified in Secs.~\ref{sec22} and \ref{sec23}, rather than as a claim of discovering the bound-state spectrum itself. The boundary conditions are purely ingoing at the event horizon and exponentially decaying at spatial infinity. The recurrence is evaluated backward and the complex roots are obtained from the vanishing of the continued-fraction residual. The calculations below use truncation order $N=500$; the residuals of the displayed roots are typically $10^{-14}$--$10^{-15}$. The numerical implementation and the data tables used for the figures are included with the source files.

At small $M\mu$, the fundamental $\ell=1$ odd-parity branch reproduces the known hydrogenic behavior
\begin{eqnarray}
\frac{\Re\omega}{\mu}&\simeq&1-\frac{(M\mu)^2}{8}, \label{eq.74}\\
-\frac{\Im\omega}{\mu}&\simeq&\frac{(M\mu)^9}{3}. \label{eq.75}
\end{eqnarray}
The continued fraction then provides the nonperturbative continuation away from the small-coupling regime.

\begin{figure*}[t]
\centering
\includegraphics[width=0.47\textwidth]{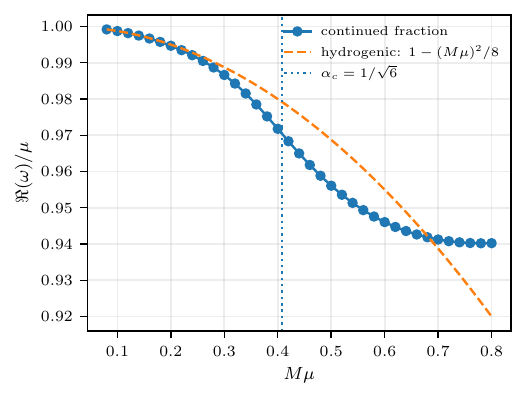}\hfill
\includegraphics[width=0.47\textwidth]{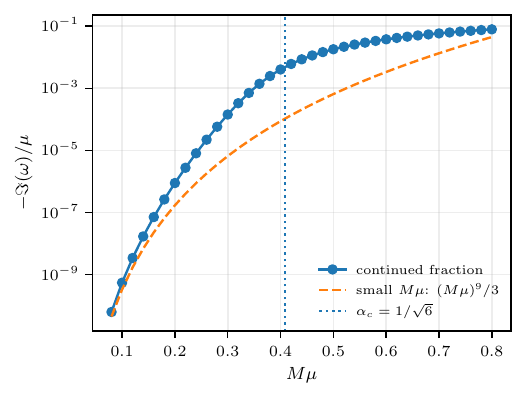}
\caption{Fundamental $\ell=1$ odd-parity Proca quasibound state. Left: real part of the frequency; the dashed curve is the leading hydrogenic approximation. Right: damping rate on a logarithmic scale; the small-coupling approximation $-\Im(\omega)/\mu\simeq(M\mu)^9/3$ is shown for reference. In both panels the vertical line marks the potential-coalescence coupling $M\mu=1/\sqrt{6}$. The pole continues across this threshold while its lifetime decreases rapidly.}
\label{fig:qbs-spectrum}
\end{figure*}

The persistence of the fundamental $\ell=1$ pole suggests that the local barrier--well coalescence need not coincide with an endpoint of the quasibound spectrum. We therefore extend the computation to $\ell=1,2,3$ and isolate the three lowest quasibound roots in the neighborhood of the corresponding critical coupling
\begin{eqnarray}
\alpha_c&=&\sqrt{\frac{\ell(\ell+1)}{12}}. \label{eq.76}
\end{eqnarray}
For each $\ell$ we label the roots $n=0,1,2$ in increasing order of $\Re\omega$ in the near-critical region. Table~\ref{tab:qbs-critical-multimode} gives their values at $\alpha=\alpha_c$.

\begin{table*}[t]
\caption{Three lowest odd-parity Proca quasibound roots at the potential-coalescence coupling $\alpha=\alpha_c$. Frequencies are expressed in units $M=1$.}
\label{tab:qbs-critical-multimode}
\begin{ruledtabular}
\begin{tabular}{cccccc}
$\ell$ & $\alpha_c$ & $n$ & $\Re(M\omega)$ & $-\Im(M\omega)$ & $\Re(\omega)/\mu$\\
\hline
1 & 0.408248 & 0 & 0.396145192 & $1.972908\times10^{-3}$ & 0.970354\\
1 & 0.408248 & 1 & 0.403244622 & $6.756492\times10^{-4}$ & 0.987744\\
1 & 0.408248 & 2 & 0.405580608 & $2.824776\times10^{-4}$ & 0.993466\\
2 & 0.707107 & 0 & 0.691726976 & $1.842593\times10^{-3}$ & 0.978250\\
2 & 0.707107 & 1 & 0.697658048 & $1.035523\times10^{-3}$ & 0.986637\\
2 & 0.707107 & 2 & 0.700770553 & $6.114024\times10^{-4}$ & 0.991039\\
3 & 1.000000 & 0 & 0.955129184 & $3.484938\times10^{-3}$ & 0.955129\\
3 & 1.000000 & 1 & 0.971609956 & $3.048939\times10^{-3}$ & 0.971610\\
3 & 1.000000 & 2 & 0.980845027 & $2.092489\times10^{-3}$ & 0.980845\\
\end{tabular}
\end{ruledtabular}
\end{table*}

\begin{figure*}[t]
\centering
\includegraphics[width=0.92\textwidth]{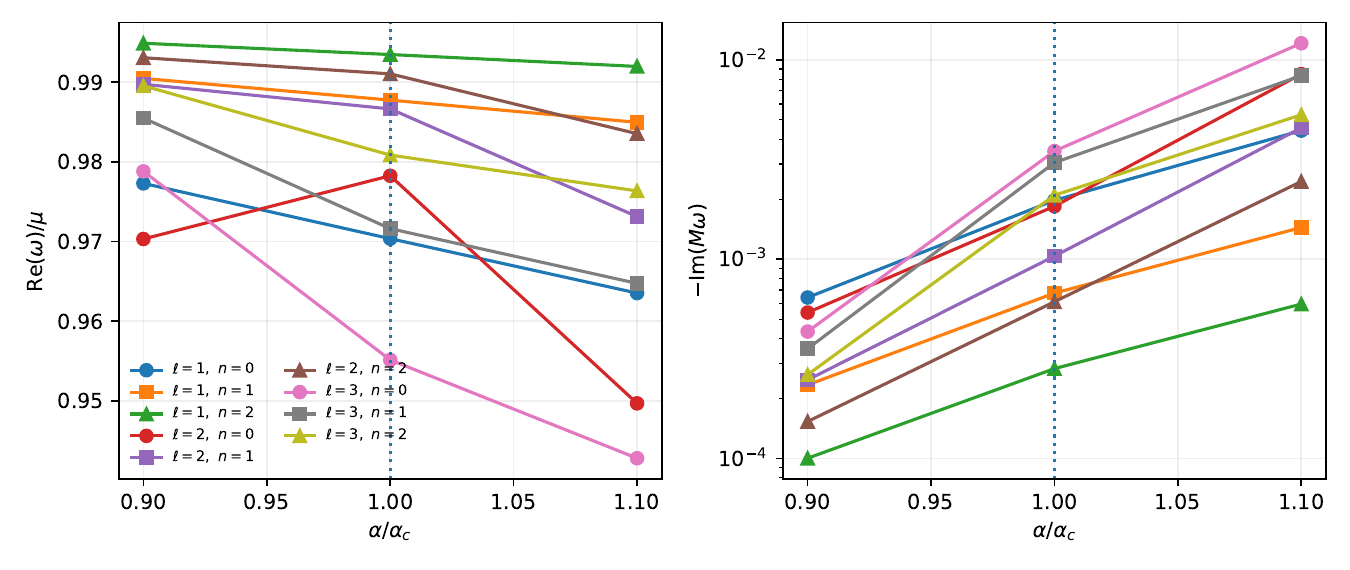}
\caption{Near-critical odd-parity Proca quasibound roots for $\ell=1,2,3$ and $n=0,1,2$. Left: $\Re(\omega)/\mu$. Right: damping rate $-\Im(M\omega)$. The vertical line marks $\alpha=\alpha_c$. The finite continuation of all displayed roots through $\alpha_c$ shows that the merger of the local potential extrema is not a termination point of the quasibound spectrum.}
\label{fig:qbs-multimode}
\end{figure*}

Figure~\ref{fig:qbs-multimode} compares the same roots at $\alpha/\alpha_c=0.9,1,1.1$. All displayed modes remain finite on both sides of the potential-coalescence threshold. The damping generally grows as the coupling is increased above criticality, while the more weakly damped roots remain closer to the mass shell $\Re\omega\simeq\mu$.

\subsection{Spectral response to geometric criticality}
\label{sec26}

The effective potential has a nonanalytic local geometry when it is parameterized by $\delta=1-\alpha^2/\alpha_c^2$: the stationary-point separation and curvature vanish as $\delta^{1/2}$, whereas the barrier--well depth vanishes as $\delta^{3/2}$. It is therefore natural to test whether the complex eigenfrequency inherits either of these exponents.

\begin{figure*}[t]
\centering
\includegraphics[width=0.92\textwidth]{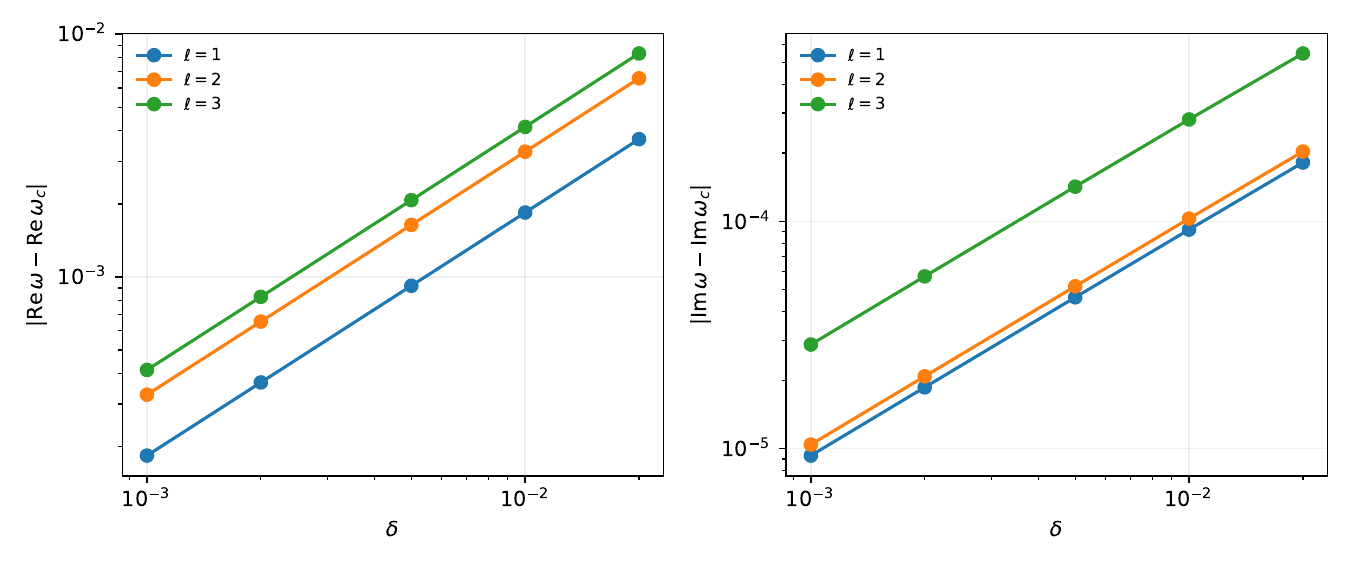}
\caption{Near-critical response of the fundamental odd-parity quasibound frequency for $\ell=1,2,3$, approached from $\alpha<\alpha_c$. Both the real and imaginary parts are consistent with an approximately linear dependence on $\delta=1-\alpha^2/\alpha_c^2$, in contrast with the nonanalytic $\delta^{1/2}$ and $\delta^{3/2}$ scaling of the local potential geometry.}
\label{fig:critical-spectral-response}
\end{figure*}

For the fundamental branch we fit the near-critical data from below to
\begin{eqnarray}
|\Re\omega-\Re\omega_c|&\propto&\delta^{p_R},\nonumber\\
|\Im\omega-\Im\omega_c|&\propto&\delta^{p_I}. \label{eq.77}
\end{eqnarray}
Using $10^{-3}\le\delta\le2\times10^{-2}$ gives
\begin{eqnarray}
(p_R,p_I)_{\ell=1}&=&(1.002,\,0.992),\nonumber\\
(p_R,p_I)_{\ell=2}&=&(1.002,\,0.992),\nonumber\\
(p_R,p_I)_{\ell=3}&=&(1.002,\,0.986). \label{eq.78}
\end{eqnarray}
Within the numerical resolution of this fit, both parts of the eigenfrequency respond approximately linearly in $\delta$. Figure~\ref{fig:critical-spectral-response} makes this behavior explicit. The spectral response is therefore smoother than the local potential geometry: the square-root merger of the extrema and the $\delta^{3/2}$ collapse of their depth do not generate a corresponding singular exponent in the fundamental complex frequency.

\subsection{Why the spectrum is regular at the fold}
\label{sec27}

The approximately unit exponents in Eq.~\eqref{eq.78} have a simple analytic interpretation. Let $\mathcal{F}(\omega,\mu^2)=0$ denote the continued-fraction spectral condition defined explicitly in Appendix~\ref{app:continued-fraction}. For a fixed multipole $\ell$, the recurrence coefficients are locally analytic functions of $\omega$ and $\mu^2$ provided the quasibound branch is kept away from the square-root threshold $\omega^2=\mu^2$ and a single branch of $q=-\sqrt{\mu^2-\omega^2}$ is chosen. The critical potential coupling does not coincide with that asymptotic branch point.

If the quasibound root at $(\omega_c,\mu_c^2)$ is simple,
\begin{eqnarray}
\left.\frac{\partial\mathcal{F}}{\partial\omega}\right|_{(\omega_c,\mu_c^2)}&\neq&0, \label{eq.79}
\end{eqnarray}
then the implicit-function theorem guarantees a locally analytic eigenfrequency $\omega(\mu^2)$. Since
\begin{eqnarray}
\mu^2-\mu_c^2&=&-\mu_c^2\,\delta, \label{eq.80}
\end{eqnarray}
we obtain generically
\begin{eqnarray}
\omega(\delta)&=&\omega_c+C_1\delta+C_2\delta^2+\mathcal{O}(\delta^3). \label{eq.81}
\end{eqnarray}
Hence the linear scaling observed numerically is the expected local behavior of a simple spectral root. By contrast, the positions of the extrema are obtained by solving a quadratic stationarity condition whose discriminant vanishes at the fold, which directly produces the square-root dependence $r_+-r_-\propto\delta^{1/2}$. The potential fold is therefore a nonanalyticity of the stationary-point parametrization, not of the radial differential operator itself.

This argument is conditional only on root simplicity and on staying away from the asymptotic threshold $\omega^2=\mu^2$. We tested root simplicity directly at the critical coupling by evaluating a centered finite-difference approximation to $\partial\mathcal{F}/\partial\omega$ for the fundamental branches. In units $M=1$, the magnitudes are approximately $91.67$, $196.67$, and $84.43$ for $\ell=1,2,3$, respectively, and are stable when the differencing step is varied from $10^{-5}$ to $3\times10^{-8}$. They are therefore manifestly nonzero for the roots considered here. The numerical continuation and the independent shooting calculation provide complementary evidence that the remaining regularity assumptions hold. In this sense, the contrast between geometric exponents $(1/2,3/2)$ and the spectral exponent $1$ follows from the analytic structure of the eigenvalue problem rather than from an accidental numerical cancellation.

\subsection{Distance from the asymptotic branch point}
\label{sec28}

A second condition entering the analyticity argument is that the mode remain separated from the massive-field branch point at $\omega^2=\mu^2$. This issue deserves an explicit check because massive-field boundary conditions are naturally formulated in terms of the complex wave number rather than $\Re\omega$ alone. Recent spectral analyses of massive Schwarzschild fields emphasize the associated two-sheeted analytic structure and the importance of fixing the relevant Riemann sheet consistently \cite{BaticEtAl2026}. For the quasibound problem we define
\begin{eqnarray}
q&=&\sqrt{\mu^2-\omega^2},\qquad \Re q>0. \label{eq.82}
\end{eqnarray}
At the fold coupling $\alpha=\alpha_c$, all nine modes displayed in Table~\ref{tab:qbs-critical-multimode} satisfy $|q|/\mu>0.11$. In particular, the smallest separation occurs for $(\ell,n)=(1,2)$ and is still finite, $|q|/\mu\simeq0.114$. Thus the geometric fold is quantitatively distinct from the asymptotic branch point for the modes used in the critical fits.

\begin{figure}[t]
\centering
\includegraphics[width=0.96\columnwidth]{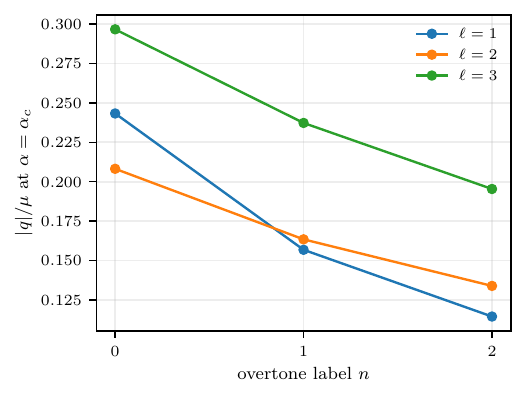}
\caption{Distance from the massive-field branch point at the potential-fold coupling. The plotted quantity $|q|/\mu$, with $q=\sqrt{\mu^2-\omega^2}$ and $\Re q>0$, remains nonzero for the first three quasibound roots at $\ell=1,2,3$. The fold therefore does not coincide with the square-root threshold of the asymptotic dispersion relation.}
\label{fig:branchpoint-distance}
\end{figure}

This diagnostic sharpens the implicit-function argument: the local nonanalyticity of the stationary radii is generated by the vanishing discriminant of the algebraic extremum equation, whereas the spectral function is evaluated at a point that is separated both from a multiple root in $\omega$ and from the asymptotic square-root singularity. The two mechanisms of nonanalyticity are therefore parametrically distinct in the present calculation.

\subsection{Independent shooting validation}
\label{sec29}

As an independent check of the continued-fraction calculation, selected fundamental roots were tested with a two-sided Riccati shooting method applied directly to the axial radial equation. Writing $y=\psi'/\psi$ with derivative with respect to $r$, one obtains
\begin{eqnarray}
\frac{\dd y}{\dd r}&=&-y^2-\frac{f'}{f}y-\frac{\omega^2-V_{\rm P}^{(-)}}{f^2}. \label{eq.83}
\end{eqnarray}
The horizon integration uses the ingoing Frobenius branch, while the outer integration uses the exponentially decaying asymptotic solution with $q=\sqrt{\mu^2-\omega^2}$ and $\Re q>0$. Matching the logarithmic derivatives provides a complex shooting residual without reference to the continued-fraction recurrence.

For the fundamental modes we repeated the shooting calculation from deliberately perturbed complex-frequency seeds rather than initializing it exactly at the continued-fraction root. At $\alpha=\alpha_c$ for $\ell=1,2,3$, and for several matching radii, the recovered frequencies agree with the continued-fraction values at the level of $7\times10^{-13}$ or better. The logarithmic-derivative mismatches remain at or below a few $10^{-11}$. We also varied the near-horizon start and matching location; the inferred frequencies remain stable at the quoted precision. This more stringent cross-check supports the smooth spectral continuation through the potential-coalescence point.

The numerical results therefore separate two notions that are easily conflated. The barrier--well pair disappears at a sharply defined coupling, but the axial quasibound poles remain regular. The critical coupling is a transition in the local topology of the effective potential and a crossover in damping, not a spectral endpoint. This conclusion is restricted to the decoupled odd-parity channel; the coupled even-parity Proca sector requires a separate matrix-valued spectral analysis.

\subsection{Boundary conditions and physical spectra}
\label{sec30}

The same radial equation can represent different physical problems depending on its boundary conditions. For quasinormal modes one imposes a purely ingoing wave at the horizon and a purely outgoing wave at infinity. Symbolically,
\begin{eqnarray}
\psi&\sim&e^{-\ii\omega r_*},\qquad r_*\rightarrow-\infty, \nonumber\\
\psi&\sim&e^{+\ii k_\infty r_*},\qquad r_*\rightarrow+\infty. \label{eq.84}
\end{eqnarray}
For Proca quasi-bound states with $\omega^2<\mu^2$, the far boundary condition is instead exponential decay,
\begin{eqnarray}
\psi&\sim&e^{-\sqrt{\mu^2-\omega^2}\,r_*},\qquad r_*\rightarrow+\infty. \label{eq.85}
\end{eqnarray}
The existence of this decaying branch is one of the clearest physical effects of the mass term.

\subsection{Dimensionless variables}
\label{sec31}

For numerical work it is convenient to scale every quantity by the black-hole mass. Define
\begin{eqnarray}
\rho&=&\frac{r}{M},\qquad \hat\omega=M\omega,\nonumber\\
\hat\mu&=&M\mu. \label{eq.86}
\end{eqnarray}
The odd Proca potential becomes
\begin{eqnarray}
M^2V_{\rm P}^{(-)}&=&\left(1-\frac{2}{\rho}\right)
\left[\frac{\ell(\ell+1)}{\rho^2}+\hat\mu^2\right]. \label{eq.87}
\end{eqnarray}
This shows that the entire test-field problem depends on the dimensionless combination $M\mu$ rather than on $M$ and $\mu$ separately.

\subsection{Structural comparison}
\label{sec32}

\begin{table*}[t]
\hbadness=10000
\centering
\caption{Structural comparison between the two spin-1 theories.}
\begin{tabular}{p{3.0cm}p{5.3cm}p{5.3cm}}
\toprule
Feature & Maxwell & Proca \\
\midrule
Mass & $\mu=0$ & $\mu\neq0$ \\
Gauge invariance & Yes & No \\
Lorenz relation & Gauge condition & Dynamical constraint \\
Physical polarizations & 2 & 3 \\
Odd sector & One master mode & One master mode \\
Even sector & One physical master mode after gauge reduction & Two coupled physical modes \\
Asymptotic potential & $0$ & $\mu^2$ \\
Radiative monopole & No & Longitudinal $\ell=0$ mode exists \\
\bottomrule
\end{tabular}
\end{table*}
\hbadness=1000

\paragraph{Flat-space radial limit.}

Set $M\rightarrow0$, so that $f\rightarrow1$ and $r_*\rightarrow r$. The Maxwell odd equation becomes
\begin{eqnarray}
\frac{\dd^2\psi}{\dd r^2}+
\left[\omega^2-\frac{\ell(\ell+1)}{r^2}\right]\psi&=&0. \label{eq.88}
\end{eqnarray}
The Proca odd equation becomes
\begin{eqnarray}
\frac{\dd^2\psi}{\dd r^2}+
\left[\omega^2-\mu^2-\frac{\ell(\ell+1)}{r^2}\right]\psi&=&0. \label{eq.89}
\end{eqnarray}
Defining
\begin{eqnarray}
k^2&=&\omega^2-\mu^2, \nonumber
\end{eqnarray}
we recover the spherical Bessel equation for a massive vector mode. The same $\ell(\ell+1)$ angular eigenvalue persists; only the radial wavenumber changes.

\paragraph{Azimuthal degeneracy.}

The Schwarzschild geometry is invariant under all rotations, not just rotations about a preferred axis. Consequently the radial equations can depend on $\ell$ but not on the azimuthal number $m$. Modes with different $m$ but the same $\ell$ are degenerate. This statement is ultimately a representation-theoretic consequence of the $SO(3)$ symmetry of the background.

The angular harmonics satisfy
\begin{eqnarray}
\mathbf{L}^2Y_{\ell m}&=&\ell(\ell+1)Y_{\ell m},\nonumber\\
L_zY_{\ell m}&=&mY_{\ell m}. \nonumber
\end{eqnarray}
The radial operator sees the Casimir eigenvalue $\ell(\ell+1)$ but no preferred projection $m$.

\section{Scalar--vector comparison}
\label{sec33}

A comparison with the Klein--Gordon field helps disentangle three ingredients that can otherwise be conflated in Schwarzschild wave equations: the spin of the field, the presence of a rest mass, and the existence or absence of gauge redundancy. For a scalar field, the separation $\Phi=e^{-i\omega t}Y_{\ell m}(\theta,\phi)u(r)/r$ leads directly to a one-dimensional equation in the tortoise coordinate. In the notation adopted here, the representative scalar and vector radial problems can therefore be compared on the same footing.

For a minimally coupled scalar field of mass $\mu_s$,
\begin{eqnarray}
\frac{\dd^2u}{\dd r_*^2}+\left[\omega^2-V_{\rm KG}(r)\right]u&=&0, \label{eq.90}\\
V_{\rm KG}(r)&=&f(r)\left[\mu_s^2+\frac{\ell(\ell+1)}{r^2}\right.\nonumber\\
&&\left.+\frac{2M}{r^3}\right]. \label{eq.91}
\end{eqnarray}
The last term is purely geometric. Even when $\ell=0$ and $\mu_s=0$, a scalar wave still feels a curvature barrier in Schwarzschild spacetime.

For the Maxwell master variable, either parity sector satisfies
\begin{eqnarray}
\frac{\dd^2\Psi_{\rm EM}}{\dd r_*^2}+\left[\omega^2-V_{\rm EM}(r)\right]\Psi_{\rm EM}&=&0, \label{eq.92}\\
V_{\rm EM}(r)&=&f(r)\nonumber\\
&&\times\frac{\ell(\ell+1)}{r^2}. \label{eq.93}
\end{eqnarray}
The absence of the scalar curvature term is not an accident. It reflects the different spin content of the perturbation. The geometry is the same, but the field samples it through a different radial operator.

For the odd-parity Proca mode,
\begin{eqnarray}
\frac{\dd^2\Psi_{\rm P}^{(-)}}{\dd r_*^2}
+\omega^2\Psi_{\rm P}^{(-)}&=&\nonumber\\
&&V_{\rm P}^{(-)}(r)\Psi_{\rm P}^{(-)}, \label{eq.94}\\
V_{\rm P}^{(-)}(r)&=&f(r)\left[\frac{\ell(\ell+1)}{r^2}+\mu_v^2\right]. \label{eq.95}
\end{eqnarray}
This is the cleanest place to see what the Proca mass does: it does not alter the angular eigenvalue, but it changes the radial dispersion and creates a nonzero asymptotic plateau. The even-parity Proca sector is richer and cannot, in general, be represented by a single potential because two physical polarizations remain coupled.

The three potentials make the division of labor between geometry and field content explicit:
\begin{eqnarray}
V_{\rm KG}-V_{\rm EM}&=&f(r)\mu_s^2\nonumber\\
&&+\frac{2Mf(r)}{r^3}, \label{eq.96}\\
V_{\rm P}^{(-)}-V_{\rm EM}
&=&f(r)\mu_v^2. \label{eq.97}
\end{eqnarray}
Thus the scalar differs from Maxwell by a spin-dependent curvature term and, if massive, by a mass term; the axial Proca mode differs from Maxwell only by the vector mass term.

\begin{figure*}[t]
\centering
\includegraphics[width=0.72\textwidth]{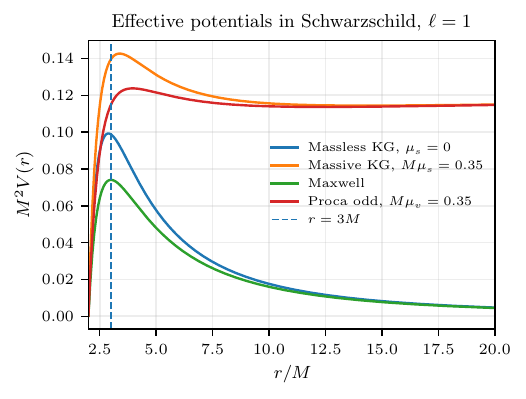}
\caption{Comparison of representative Schwarzschild effective potentials for a massless scalar field, a massive scalar field, the Maxwell field, and the axial Proca mode. The scalar curvature term changes the barrier near the black hole, whereas a mass term changes the asymptotic level. Parameters: $M=1$, $\ell=1$, and $\mu_s=\mu_v=0.35$ for the massive curves.}
\label{fig:scalar-vector-bridge}
\end{figure*}

Figure~\ref{fig:scalar-vector-bridge} displays these statements graphically for one representative multipole. The plot is not meant to suggest that the even Proca sector has been reduced to a single master potential; it compares only the decoupled axial vector mode with the scalar and electromagnetic cases.

\subsection{Angular spectrum and field content}
\label{sec34}

All three theories are organized by the same spherical symmetry. The scalar field uses ordinary spherical harmonics,
\begin{eqnarray}
- D^AD_A Y_{\ell m}&=&\ell(\ell+1)Y_{\ell m}, \label{eq.98}
\end{eqnarray}
whereas the vector theories use the polar and axial vector harmonics constructed from $Y_{\ell m}$. In this sense, the number $\ell(\ell+1)$ remains the angular backbone of the problem. What changes is the number of independent radial amplitudes and the way they are constrained.

For a real Klein--Gordon field there is one local physical degree of freedom. Maxwell theory has two physical transverse polarizations because the four-potential contains gauge redundancy. Proca theory has three physical polarizations: two transverse and one longitudinal. A compact way to remember the counting is
\begin{eqnarray}
N_{\rm dof}^{\rm KG}&=&1,\nonumber\\
N_{\rm dof}^{\rm Maxwell}&=&2,\nonumber\\
N_{\rm dof}^{\rm Proca}&=&3. \label{eq.99}
\end{eqnarray}
The increase from two to three in the vector case is precisely why the massive even-parity system does not collapse to the Maxwell result by a naive substitution.

\subsection{Asymptotic dispersion}
\label{sec35}

The asymptotic region also provides a useful comparison. Far from the black hole, $f(r)\rightarrow1$ and the centrifugal terms vanish. The scalar and Proca fields then have the same massive dispersion relation,
\begin{eqnarray}
\omega^2&=&k^2+\mu^2, \label{eq.100}
\end{eqnarray}
while the Maxwell field obeys
\begin{eqnarray}
\omega^2&=&k^2. \label{eq.101}
\end{eqnarray}
The equality of the scalar and Proca dispersion laws does not imply equality of the theories. It only says that the energy--momentum relation of a free relativistic excitation depends on its mass, not on whether that excitation carries spin zero or spin one. Their polarization structure and their coupling to curvature remain different.

\begin{table*}[t]
\hbadness=10000
\centering
\caption{Scalar and vector fields in Schwarzschild spacetime: a compact comparison.}
\label{tab:kg-maxwell-proca}
\begin{tabular}{p{3.0cm}p{3.8cm}p{3.8cm}p{3.8cm}}
\toprule
Feature & Klein--Gordon & Maxwell & Proca\\
\midrule
Spin & $0$ & $1$ & $1$\\
Gauge symmetry & none & $U(1)$ & absent for the mass term used here\\
Physical degrees of freedom & $1$ & $2$ & $3$\\
Angular building blocks & $Y_{\ell m}$ & vector harmonics from $Y_{\ell m}$ & vector harmonics from $Y_{\ell m}$\\
Radiative multipoles & $\ell\geq0$ & $\ell\geq1$; $\ell=0$ is Coulombic & $\ell\geq0$, including a longitudinal monopole\\
Asymptotic potential & $\mu_s^2$ & $0$ & $\mu_v^2$ for the odd mode\\
Representative Schwarzschild potential & $f[\mu_s^2+\ell(\ell+1)/r^2+2M/r^3]$ & $f\ell(\ell+1)/r^2$ & $f[\ell(\ell+1)/r^2+\mu_v^2]$ (odd)\\
\bottomrule
\end{tabular}
\end{table*}
\hbadness=1000

The comparison makes a useful separation of roles. The repeated appearance of $\ell(\ell+1)$ reflects the spectrum of the two-sphere, whereas the rest of the radial operator retains information about the field representation: spin, mass, constraints, and gauge structure are not encoded by the angular eigenvalue alone.

\section{Discussion}
\label{sec:discussion}
The distinction emerging from this calculation concerns the mathematical objects whose criticality is being assessed. A repeated solution of $\partial_r V_{\rm ax}=0$ is an algebraic event in a real, local function. A multiple quasibound eigenfrequency, by contrast, requires a repeated zero of a complex global spectral condition. In general it would require $F(\omega,\delta)=0$ and $\partial_\omega F(\omega,\delta)=0$, with additional eigenfunction structure needed before identifying an exceptional point. No such coincidence follows from the stationary-point discriminant.

The axial potential maps, after division by $\mu^2$, onto the standard timelike Schwarzschild geodesic potential with $\mathcal L_{\rm eff}^2=\ell(\ell+1)/\mu^2$. This observation makes the critical radius $6M$ physically intelligible as the image of the known ISCO stability boundary~\cite{Darwin1959,Chandrasekhar}. It is not a new characteristic Schwarzschild radius, nor is it evidence that the Proca wave equation reduces to the particle geodesic equation. The field-theoretic content lies in comparing the exact potential fold with the response of a complex boundary-value spectrum.

Rosa and Dolan~\cite{RosaDolan} already established the odd-parity recurrence and quasibound branches, while Ge, Matsumoto and Zhang~\cite{GeMatsumotoZhang} have approached the same spectral problem using Seiberg--Witten/quasinormal-mode duality. Our numerical results are best interpreted as a targeted continuation study, not as new Proca eigenvalue families. We use multiple multipoles and overtones to demonstrate finite mode continuation through the potential fold and an independent Riccati shooting test to reduce reliance on one numerical representation. Small continued-fraction residuals indicate consistency of the truncated spectral equation but must not be confused with rigorous eigenfrequency error bounds.

The exact relation $U(x,\delta)=U_c(x)-[\ell(\ell+1)/12](1-2/x)\delta$ explains the contrast. The potential at each fixed $x>2$ is analytic, indeed affine, in the fold parameter, although the positions obtained by solving $\partial_x U=0$ acquire $\sqrt\delta$. Analytic dependence of the operator is not by itself sufficient to prove analytic dependence of all spectral roots: this also requires a suitable analytic continuation of the boundary conditions, an isolated simple zero of the spectral function, and separation from $\omega^2=\mu^2$. These conditions are investigated numerically for the reported branches, and the resulting approximately linear exponents are consistent with the conditional implicit-function argument. The conclusion does not exclude multiple roots or nonanalytic behavior in other portions of the spectrum.

A useful comparison is provided by scalar-field studies that locate changes in quasinormal spectra directly~\cite{AlvesPonquioMedeiros} and that track complex wavenumber sheets through a massive-field threshold~\cite{BaticEtAl2026}. The parameters and spectral boundary conditions differ from the potential-defined fold considered here. Moreover, the observation that a secondary potential well need not fix black-hole ringdown spectral stability~\cite{WangWu2026} supports a broader methodological warning, without implying equivalence between the models. What is new in our analysis is the \emph{joint quantitative comparison} of the spin-one potential-fold exponents, the geodesic identification and selected complex quasibound continuations within one explicitly controlled Schwarzschild problem.

\subsection{Independent numerical diagnostics of spectral-root simplicity}
\label{sec:independent_diagnostics}
The distinction between a stationary-point fold and a multiple spectral root can be tested directly with the same continued-fraction function $F(\omega,\mu^2)$ used to locate the quasibound frequencies. We performed a fresh continuation of the nine modes at their respective fold masses using two separate truncation orders, $N=500$ and $N=800$, and the same ingoing/decaying branch conventions as in the main calculation. All nine root searches converged at both truncations. At the reported numerical precision the maximum difference $|\omega_{800}-\omega_{500}|$ was zero, while the largest absolute continued-fraction residual at $N=800$ was $4.54\times 10^{-14}$. This check demonstrates numerical consistency between these truncations; it is not a certified error bound, since systematic errors common to both may remain.

For the fundamental branch, we also evaluated a symmetric finite-difference derivative of the spectral residual at the fold, varying its step size. Table~\ref{tab:root_simplicity} records the representative magnitudes at step $h=10^{-7}$ and the maximum differences with independently obtained two-sided Riccati-shooting roots found from perturbed initial seeds and different matching radii. The derivative magnitudes are clearly separated from zero for the branches displayed, giving direct numerical evidence against a multiple zero at these specific frequencies. Their absolute values depend on the normalization of the chosen spectral residual; only the property of being nonzero is invariant under multiplication of $F$ by a locally nonvanishing analytic function.

\begin{table}[t]
\centering
\caption{Numerical diagnostics at the respective axial potential-fold couplings. $|\partial_\omega F|$ is evaluated by a centered difference at $h=10^{-7}$; $\Delta_{\rm shoot}$ is the largest difference between the fundamental continued-fraction root and the independently recovered shooting roots in the supplied robustness sweep. The values are numerical diagnostics, not rigorous enclosure bounds.}
\label{tab:root_simplicity}
\begin{tabular}{cccc}
\toprule
$\ell$ & $|\partial_\omega F|$ & $\Delta_{\rm shoot}$ & $\min_n |q_n|/\mu$\\
\midrule
$1$ & $91.6687$ & $1.05\times10^{-13}$ & $0.114$\\
$2$ & $196.6681$ & $1.95\times10^{-13}$ & $0.134$\\
$3$ & $84.4311$ & $6.93\times10^{-13}$ & $0.195$\\
\bottomrule
\end{tabular}
\end{table}

The last column also tests separation from the branch point $q=\sqrt{\mu^2-\omega^2}=0$ for the three tracked overtones at each $\ell$. The original numerical CSV files and a short independent recheck script accompany the sources, permitting the reader to reproduce the table's continued-fraction comparisons without changing the field equations or the underlying spectral recurrence. Crucially, these results support simple-root behavior only for the sampled roots and selected parameter values, not a global absence of exceptional points.

A recent mathematical analysis by Gunara~\cite{Gunara2026} addresses massive Maxwell--Proca threshold asymptotics and quasibound resonances on subextremal Reissner--Nordstr\"om exteriors, encompassing Schwarzschild as the uncharged limit. Its focus is the massive spectral threshold, late-time decay and polarization-resolved resolvent structure; the present analysis instead quantifies a finite-radius stationary-point fold of the decoupled Schwarzschild axial potential and tests selected nearby quasibound roots. These results should be treated as complementary rather than as competing discoveries of the same spectrum.

The coupled parity-even Proca modes are not described by a single axial Schr\"odinger potential and their spectral condition is matrix-valued~\cite{RosaDolan}. Thus our conclusions must not be extrapolated to the entire massive-vector spectrum. A natural extension is to investigate whether matrix-valued polar spectral roots can become multiple at dynamically distinct couplings, and to test the corresponding eigenfunctions and numerical conditioning. Such an extension would require a separate calculation rather than an inference from axial potential geometry.

\section{Conclusions}
\label{sec:conclusions}
We have compared Maxwell and massive Proca fields in Schwarzschild spacetime, focusing on whether coalescence of the two exterior stationary points of the odd-parity Proca potential must be accompanied by a singular or degenerate quasibound spectrum. For the specific axial modes tracked in this work, the answer is negative.

The effective potential has an exact formal correspondence with the timelike Schwarzschild-geodesic radial potential. As the dimensionless Proca mass approaches $(M\mu_c)^2=\ell(\ell+1)/12$ from below, its exterior maximum and minimum merge at the standard ISCO radius $r=6M$. With $\delta=1-(M\mu)^2/(M\mu_c)^2$, we obtain analytically the three different critical behaviors: stationary-point separation proportional to $\delta^{1/2}$, barrier--well depth proportional to $\delta^{3/2}$, and stationary-point curvatures proportional to $\delta^{1/2}$. The $6M$ scale and the geodesic potential are established results; their role here is to identify and classify the axial potential fold exactly.

For $\ell=1,2,3$ and the first three reported quasibound branches, numerical continued fractions show finite continuation on both sides of the fold. Fits to the fundamental complex frequencies give a response close to linear in $\delta$, unlike the nonanalytic displacement of the potential extrema. Selected roots agree with a separate Riccati shooting calculation. This contrast is consistent with the analytic dependence of the fixed-radius radial operator and the implicit-function theorem applied to a simple spectral zero on a fixed asymptotic sheet.

Our result is therefore a distinction between local geometrical criticality and the global complex spectrum, rather than a claim of a new Proca equation, a newly discovered quasibound spectrum, or a universal theorem about all massive-vector modes. Extensions to coupled even-parity modes, possible spectral exceptional points and certified global numerical error bounds remain meaningful questions for future investigation.

\appendix
\section{Explicit tortoise coordinate}
\label{sec44}

From Eq.~\eqref{eq.28},
\begin{eqnarray}
\frac{\dd r_*}{\dd r}&=&\frac{1}{1-2M/r}=\frac{r}{r-2M}. \nonumber
\end{eqnarray}
Write
\begin{eqnarray}
\frac{r}{r-2M}&=&1+\frac{2M}{r-2M}. \nonumber
\end{eqnarray}
Integration gives
\begin{eqnarray}
r_*&=&r+2M\ln\left|\frac{r}{2M}-1\right|+C. \label{eq.108}
\end{eqnarray}
The additive constant is physically irrelevant. As $r\rightarrow2M^+$, $r_*\rightarrow-\infty$.

\section{Restoring ordinary units}
\label{sec45}

The Schwarzschild function is
\begin{eqnarray}
f(r)&=&1-\frac{2GM}{c^2r}. \nonumber
\end{eqnarray}
If the vector particle has physical mass $m_v$, the geometric mass parameter is
\begin{eqnarray}
\mu&=&\frac{m_vc}{\hbar}. \label{eq.109}
\end{eqnarray}
The dimensionless coupling that controls the bound-state problem is then
\begin{eqnarray}
M\mu&=&\frac{GMm_v}{\hbar c}. \label{eq.110}
\end{eqnarray}
This quantity compares the gravitational radius of the black hole with the Compton wavelength of the vector field.

\section{Two-sphere identities}
\label{sec46}

The scalar harmonics are normalized by
\begin{eqnarray}
\int_{S^2}Y_{\ell m}^*Y_{\ell' m'}\,\dd\Omega
&=&\delta_{\ell\ell'}\delta_{mm'}. \label{eq.111}
\end{eqnarray}
Using integration by parts and Eq.~\eqref{eq.21},
\begin{eqnarray}
\int_{S^2}Y_A^{\ell m*}Y^{A\,\ell' m'}\,\dd\Omega
&=&\ell(\ell+1)\delta_{\ell\ell'}\delta_{mm'}, \label{eq.112}
\end{eqnarray}
and similarly
\begin{eqnarray}
\int_{S^2}X_A^{\ell m*}X^{A\,\ell' m'}\,\dd\Omega
&=&\ell(\ell+1)\delta_{\ell\ell'}\delta_{mm'}. \label{eq.113}
\end{eqnarray}
The polar and axial harmonics are mutually orthogonal:
\begin{eqnarray}
\int_{S^2}Y_A^{\ell m*}X^{A\,\ell' m'}\,\dd\Omega&=&0. \label{eq.114}
\end{eqnarray}
These relations are what allow the field equations to be projected mode by mode.

\section{Odd-parity continued-fraction equation}
\label{sec47}
\label{app:continued-fraction}

For completeness, the numerical quasibound-state calculation in Sec.~\ref{sec25} uses the standard Frobenius ansatz for the decoupled odd-parity Proca equation \cite{RosaDolan},
\begin{eqnarray}
\psi(r)&=&f(r)^{-2\ii\omega}r^{-\nu}e^{qr}
\sum_{n=0}^{\infty}a_n[f(r)]^n, \label{eq.115}\\
q&=&-\sqrt{\mu^2-\omega^2},\qquad
\nu=\frac{\omega^2-q^2}{q}. \label{eq.116}
\end{eqnarray}
The square-root branch is chosen so that $\Re\sqrt{\mu^2-\omega^2}>0$. The coefficients satisfy
\begin{eqnarray}
\alpha_0a_1+\beta_0a_0&=&0,\nonumber\\
\alpha_na_{n+1}+\beta_na_n+\gamma_na_{n-1}&=&0,\qquad n>0, \label{eq.117}
\end{eqnarray}
with
\begin{eqnarray}
\alpha_n&=&(n+1)(n+1-4\ii\omega), \label{eq.118}\\
\beta_n&=&-2n^2+\frac{2}{q}(4\ii\omega q+3q^2-\omega^2-q)n\nonumber\\
&&+\frac{1}{q}\left[3q^2+(12q-1)\omega^2-4q^3\right.\nonumber\\
&&\left.\qquad-q\ell(\ell+1)-12\ii\omega q^2\right.\nonumber\\
&&\left.\qquad+4\ii\omega q+4\ii\omega^3\right], \label{eq.119}\\
\gamma_n&=&n^2-\frac{2}{q}(q^2-\omega^2+2\ii\omega q)n\nonumber\\
&&-\frac{1}{q^2}\left[-4\ii\omega q^3+4\ii\omega^3q\right.\nonumber\\
&&\left.\qquad+6\omega^2q^2-q^4-\omega^4+q^2\right]. \label{eq.120}
\end{eqnarray}
The minimal solution is obtained from the infinite continued fraction. Numerically, we truncate at $N$ and evaluate the denominator backward,
\begin{eqnarray}
D_N&=&\beta_N,\nonumber\\
D_n&=&\beta_n-\frac{\alpha_n\gamma_{n+1}}{D_{n+1}},\qquad n=N-1,\ldots,1,\nonumber\\
0&=&\beta_0-\frac{\alpha_0\gamma_1}{D_1}. \label{eq.121}
\end{eqnarray}
The last equation is solved in the complex $\omega$ plane. The calculations reported in the main text use $N=500$; the resulting roots are additionally checked against the independent two-sided Riccati shooting construction described in Sec.~\ref{sec29}.

\section*{Data availability}
The numerical data supporting the findings of this study, including the computed quasibound-state frequencies, and the computational scripts used for the continued-fraction and Riccati-shooting calculations are available from the corresponding author upon reasonable request. No experimental or observational datasets were used in this study.

\section*{Use of AI tools}
During the preparation of this manuscript, the authors used ChatGPT (OpenAI) to assist with English-language editing, manuscript organization, LaTeX formatting, bibliographic consistency checks, and the refinement of schematic illustrations. The Python scripts used in this study were developed by the authors, with AI assistance limited to code adjustments, debugging, and troubleshooting. All AI-assisted material was critically reviewed and verified by the authors, who retain full responsibility for the mathematical derivations, numerical results, scientific interpretations, references, and conclusions presented in this work.

\hbadness=10000

\end{document}